\documentclass[aps,reprint,superscriptaddress,longbibliography,floatfix]{revtex4-2}

\PassOptionsToPackage{hidelinks}{hyperref}

\usepackage{amsmath,amssymb,amsfonts}
\usepackage[T1]{fontenc}

\usepackage{graphicx}
\graphicspath{{Figures/}}
\usepackage[caption=false]{subfig}

\usepackage{lineno}

\usepackage{eurosym}

\usepackage{quantikz}
\usepackage{orcidlink}

\usepackage{hyperref}

\newcommand{\bitem}{\begin{itemize}}
\newcommand{\fitem}{\end{itemize}}
\newcommand{\beq}{\begin{equation}}
\newcommand{\eeq}{\end{equation}}
\newcommand{\beqa}{\begin{eqnarray}}
\newcommand{\eeqa}{\end{eqnarray}}

\begin{document}

\author{Lucas Ferreira R. de Moura}
\affiliation{Instituto de F\'{i}sica, Universidade Federal de Goi\'{a}s, 74.001-970, Goi\^{a}nia - GO, Brazil}

\author{Daniel Y. Akamatsu}
\affiliation{Instituto de F\'{i}sica, Universidade Federal de Goi\'{a}s, 74.001-970, Goi\^{a}nia - GO, Brazil}

\author{G. D. de Moraes Neto~\orcidlink{0000-0003-4273-8380}}
\email{gdmneto@gmail.com}
\affiliation{Faculty of Civil Engineering and Mechanics, Kunming University of Science and Technology, Kunming, 650500, China}

\author{Norton G. de Almeida~\orcidlink{0000-0001-8517-6774}}
\affiliation{Instituto de F\'{i}sica, Universidade Federal de Goi\'{a}s, 74.001-970, Goi\^{a}nia - GO, Brazil}
\affiliation{Instituto de F\'{i}sica de S\~{a}o Carlos, Universidade de S\~{a}o Paulo, Caixa Postal 369, 13560-970, S\~{a}o Carlos - SP, Brazil.}

\title{Visibility-Engineered Multiparameter Sensing in Dispersive Quantum Interferometry: Identifiability, Noise Robustness, and Hardware Emulation}


\begin{abstract}
	We propose a dispersive interferometric protocol for simultaneously estimating the dimensionless inverse temperature  $b=\beta\hbar\omega_0$ of a thermal two-level ancilla and the dispersive phase $x=\chi t/2$.
	Photon counting at the interferometer output yields probabilities governed by the fringe visibility $\mathcal{V}(b)$; its magnitude controls phase sensitivity, while its derivative sets thermometric sensitivity.
	A single binary outcome provides only a rank-one Fisher matrix, so local identifiability requires combining several controlled reference phases. We build a multi-setting likelihood and numerically demonstrate that a maximum-likelihood estimator attains the Cram\'er--Rao bound asymptotically.
	For $N=1$ we implement the equivalent qubit circuit on IBM Quantum hardware, interpreting the data as a conservative likelihood emulation; the observed contrast shrinkage is consistent with a noise bias toward infinite temperature.
	We also compare NOON, cat, and squeezed probe states under amplitude and phase damping, highlighting windows of usable joint sensitivity. The framework clarifies what can be inferred from photon counting statistics and which control settings are necessary for genuine multiparameter estimation in realistic platforms.
\end{abstract}

\maketitle

\section{Introduction}
\label{sec:intro}

Quantum metrology harnesses quantum resources---such as coherence, entanglement, and nonclassical light---to achieve a fundamental enhancement in the precision of estimating physical parameters, surpassing classical limits~\cite{giovannetti2004quantum,giovannetti2006quantum,giovannetti2011advances,paris2009quantum,xiang2013,toth2014quantum,degen2017quantum,montenegro2025}. In this framework, the ultimate precision bounds are governed by the classical Fisher information (FI) and its quantum generalization, the quantum Fisher information (QFI), which appear in the Cram\'{e}r--Rao inequalities for unbiased estimators~\cite{cramer1999mathematical,LeCam-1986,vantrees1968detection,helstrom1967minimum,helstrom1969quantum,braunstein1994statistical,paris2009quantum,sidhu2020geometric}. When multiple parameters are estimated simultaneously, these bounds are dictated by the full Fisher-information matrix. In this multiparameter setting, non-trivial trade-offs can arise from statistical correlations and from the potential incompatibility of the optimal measurements for different parameters~\cite{helstrom1976quantum, knott2016local, sidhu2020geometric, roccia2017entangling, liu2020quantum, Mihailescu2024, xia2023toward, yang2025overcoming}.

Multiparameter quantum metrology is particularly relevant in scenarios where a single experimental platform must be calibrated or characterized with respect to several unknown quantities at once. It has attracted considerable attention because it enables the simultaneous extraction of information about several parameters within a single experimental protocol, although the attainable advantage over independent estimation strategies depends on the compatibility of the parameters and the measurement scheme\cite{szczykulska2016multi,
humphreys2013quantum,
ragy2016compatibility,gessner2018sensitivity,liu2020quantum}. Representative examples include the simultaneous estimation of phase and loss in interferometry~\cite{ou1997fundamental,giovannetti2004quantum,giovannetti2006quantum,yadav2024quantum,grond2011mach}, and quantum thermometry where temperature must be inferred alongside unknown Hamiltonian parameters~\cite{mehboudi2019thermometry,depasquale2018,correa2015individual,campbell2017,rubio2021,mok2021optimal,glatthard2022optimal,mehboudi2022fundamental,mihailescu2023thermometry,Mihailescu2024,sekatski2022,planella2022,plodzien2018few,aybar2022criticalquantum,hovhannisyan2021optimal,jorgensen2022bayesian,ullah2025optimizing,ullah2025optimal,ostermann2023temperature}. Experimentally, nanoscale and solid-state quantum thermometry have been realized using luminescent nanothermometers, color centers in diamond, and circuit- or optomechanical platforms~\cite{brites2012,djavcanin2023luminescence,quintanilla2022challenges,fujiwara2021diamond,kucsko2013nanometre,scigliuzzo2020primary,montenegro2020}. 

Optical interferometers, most notably the Mach--Zehnder interferometer (MZI), serve as a natural and flexible testbed for this task~\cite{ou1997fundamental, pezze2007phase, grond2011mach, gatto2022heisenberg, kumar2023enhanced, yadav2024quantum, abdellaoui2024quantum, verma2025advantage,rai2025heisenberg}. Closely related nonlinear interferometric architectures, such as SU(1,1) interferometers, have also attracted considerable interest owing to their enhanced phase sensitivity and quantum-limited performance under a variety of input states and detection schemes~\cite{chang2022improvement,liu2023enhancement,abouelkhir2025enhancing,jiang2025phase}. Key examples include quantum imaging and sensing~\cite{dowling2015quantum,degen2017quantum,montenegro2025,white1998interaction, yan2016survey, ahmadi2023high, hance2021counterfactual, wang2022review, yang2023extractable, yang2024quantum}. They allow parameters to be encoded into relative phases or via dispersive light--matter interactions and can operate with a wide variety of quantum input states. Concurrently, the field of digital quantum simulation enables the implementation of these schemes as quantum circuits, where beam splitters and interactions are mapped onto sequences of unitary gates acting on qubits~\cite{Preskill2018, martinis2015qubit, ma2021adaptive, bharti2022noisy, devoret2013superconducting, blais2021circuit, qiskit, ibmq, Beckey2020, Kaubruegger2019, Meyer2020, akamatsu2025fundamental}.

For a quantum-technology implementation, however, it is not sufficient to display large Fisher-information elements.  A deployable multiparameter sensor must specify the measurement settings, the estimator applied to the raw count record, the local identifiability conditions, and the noise model that determines whether the ideal information can be accessed.  This practical perspective motivates the central organization of the present work: we separate single-setting sensitivity diagnostics from a full multi-setting estimation protocol, and we treat noisy hardware execution as an emulation and calibration tool unless estimator-level error scaling is demonstrated.

In this work, we investigate a model in which an optical field interacts dispersively with a thermalized two-level system placed in one arm of an MZI~\cite{stace2010quantum}. The Hamiltonian governing the interaction is
\begin{equation}
    \hat{H} = \hbar \chi \, \hat{a}^\dagger \hat{a} \, \hat{\sigma}_+ \hat{\sigma}_-,
    \label{eq:dispersive_H}
\end{equation}
leading to the evolution operator
\begin{equation}
    \hat{U} = e^{-i 2x \hat{a}^\dagger \hat{a} \, \hat{\sigma}_+ \hat{\sigma}_-}, \qquad x = \chi t/2,
    \label{eq:dispersive_U}
\end{equation}
where $t$ is the interaction time, $\chi$ the dispersive coupling strength, $x$ the resulting dimensionless half-phase coordinate, $\hat{a}$ the optical annihilation operator, and $\hat{\sigma}_+\hat{\sigma}_- = \ket{e}\!\bra{e}$ the projector onto the atomic excited state. Such qubit- or impurity-based probes have been extensively studied in quantum thermometry and dephasing-based sensing~\cite{razavian2019,brunelli2012qubit,martin2013berry,sabin2014,campbell2018precision}.

This study extends our recent single-parameter analysis of dispersive quantum thermometry~\cite{akamatsu2025fundamental} to a measurement-specific multiparameter setting.  In that prior work, we showed that the Hamiltonian parameter $x$ can profit from photon-number enhancement whereas temperature estimation is constrained by the thermal response of the ancilla.  Here we keep the same physical mechanism but formulate it in terms of the classical likelihood generated by photon counting.  The useful signal is governed by a single operational quantity, the thermal visibility $\mathcal{V}(\beta)$, which controls the fringe contrast, and by the derivative $\mathcal{V}'(\beta)$, which controls how strongly the likelihood changes with temperature.

Because the photon-counting probabilities are entirely determined by the thermal visibility $\mathcal{V}(\beta)$, the corresponding single-setting Fisher-information landscapes are naturally interpreted as \emph{visibility-based sensitivity maps}. These maps quantify the local sensitivity of the measurement outcomes to variations of $(\beta,x)$ but, as discussed below, do not by themselves constitute an operational multiparameter estimation protocol.

A central point is the local-identifiability constraint.  A single photon-counting setting has only two outcomes, so its Fisher-information matrix for the two parameters $(\beta,x)$ is rank one.  It is therefore useful as a diagnostic information landscape, or for estimating one parameter when the other is calibrated, but it is not by itself a full local two-parameter estimator.  We consequently formulate simultaneous estimation through a total Fisher matrix $F_{\rm tot}=\sum_j F^{(j)}$ built from several controlled settings, for example different reference phases or interrogation configurations.  This resolves the distinction between visibility-based information diagnostics and a genuinely invertible multiparameter Cram\'er--Rao bound.

Moving beyond the ideal unitary limit, we analyze how amplitude damping and phase damping degrade the photon-counting information.  Rather than claiming a universal probe-family robustness comparison from a small set of examples, we present a comparative robustness study of three standard probe families---NOON, cat, and squeezed states---under matched noise assumptions.  The resulting message is conservative: NOON probes give large ideal phase response but are fragile to loss, whereas cat and squeezed probes can retain usable information over broader noise windows depending on the operating regime.

Complementing the analytical treatment, we construct a quantum circuit that reproduces the $N=1$ likelihood model and execute it on IBM Quantum hardware.  These runs are described as hardware emulation of the interferometric statistics, not as a full metrological validation.  A full validation would require blind or pseudo-blind parameter estimation, empirical mean-square errors, and comparison with the appropriate Cram\'er--Rao bound as a function of resources.  The present hardware data instead serve to test the likelihood reconstruction and to identify systematic contrast shrinkage caused by device noise and state-preparation-and-measurement errors.

Unlike our previous single-parameter thermometric analysis, the present work addresses the local-identifiability problem associated with simultaneous estimation of temperature and dispersive phase. The central contribution is not merely the calculation of additional Fisher-information elements, but the demonstration that binary photon-counting statistics are intrinsically rank deficient for two-parameter estimation and therefore require a multi-setting protocol. In addition, the NOON-state excitation number $N$ determines the periodicity and aliasing scale of the likelihood through $\cos(2Nx)$, providing a direct control parameter for selecting a locally unambiguous estimation window. In this framework, the thermal visibility emerges as a unifying quantity governing both thermometric and phase sensitivities.

The paper is organized as follows. Section~\ref{sec:theory} reviews classical and quantum Fisher-information tools and defines the operational cost used for joint estimation. Section~\ref{sec:model} presents the interferometric likelihood, the visibility-controlled Fisher elements, and the multi-setting Fisher matrix needed for full local identifiability. Section~\ref{sec:circuit} details the digital circuit implementation. Section~\ref{sec:experiment_ibm} presents hardware emulation on IBM Quantum hardware. Section~\ref{sec:robustness_analysis} discusses probe-family robustness under amplitude and phase damping and interprets the hardware deviations. Finally, Sec.~\ref{sec:conclusions} summarizes the main operational conclusions and the requirements for a complete sensing experiment.

\section{Quantum multiparameter estimation}
\label{sec:theory}

In this section we briefly review the basic tools of classical and quantum multiparameter estimation that will be used throughout the paper.
We follow standard treatments of quantum metrology and quantum estimation theory, adapting the notation to our specific two-parameter problem.

\subsection{Classical Fisher information}

Let $\boldsymbol{\theta} = (\theta_1,\theta_2,\dots,\theta_p)^\top$ denote a vector of $p$ real parameters to be estimated from the outcomes $y$ of a measurement described by a conditional probability distribution $p(y|\boldsymbol{\theta})$.
An estimator $\hat{\boldsymbol{\theta}}(y)$ is a function of the observed data, and its performance is characterized by the covariance matrix
\begin{equation}
    \mathrm{Cov}(\hat{\boldsymbol{\theta}})_{ab}
    = \big\langle (\hat{\theta}_a - \theta_a) (\hat{\theta}_b - \theta_b) \big\rangle,
\end{equation}
where the average is taken with respect to $p(y|\boldsymbol{\theta})$.
For unbiased estimators, $\langle \hat{\theta}_a \rangle = \theta_a$, the covariance matrix is bounded from below by the inverse of the classical Fisher-information matrix (FIM) $F(\boldsymbol{\theta})$, whose elements are defined as
\begin{equation}
    F_{ab}(\boldsymbol{\theta})
    = \sum_y \frac{1}{p(y|\boldsymbol{\theta})}
      \frac{\partial p(y|\boldsymbol{\theta})}{\partial \theta_a}
      \frac{\partial p(y|\boldsymbol{\theta})}{\partial \theta_b},
    \label{eq:classical_FIM_def}
\end{equation}
for discrete outcomes $y$.
In matrix form, the multiparameter Cram\'{e}r--Rao bound (CRB) reads
\begin{equation}
    \mathrm{Cov}(\hat{\boldsymbol{\theta}})
    \;\succeq\; \frac{1}{\mu}\,F(\boldsymbol{\theta})^{-1},
    \label{eq:CRB_classical}
\end{equation}
where $\mu$ is the number of independent repetitions of the experiment and $A \succeq B$ means that $A-B$ is a positive semidefinite matrix.
In particular, for any real weight vector $\boldsymbol{c}$ one has
\begin{equation}
    \boldsymbol{c}^\top \mathrm{Cov}(\hat{\boldsymbol{\theta}})\,\boldsymbol{c}
    \;\ge\; \frac{1}{\mu}\,
    \boldsymbol{c}^\top F(\boldsymbol{\theta})^{-1} \boldsymbol{c},
\end{equation}
showing that the FIM encapsulates all attainable precisions for unbiased estimators built from the chosen measurement statistics.

\subsection{Quantum Fisher-information matrix and quantum Cram\'{e}r--Rao bound}

In the quantum setting, the parameters $\boldsymbol{\theta}$ are encoded in a family of density operators $\hat{\rho}(\boldsymbol{\theta})$ acting on a Hilbert space $\mathcal{H}$.
A general measurement is described by a positive operator-valued measure (POVM) $\{\hat{\Pi}_y\}$, with $\hat{\Pi}_y\ge 0$ and $\sum_y \hat{\Pi}_y = \mathbb{I}$, leading to the Born probabilities
\begin{equation}
    p(y|\boldsymbol{\theta}) = \mathrm{Tr}\big[\hat{\rho}(\boldsymbol{\theta}) \hat{\Pi}_y\big].
    \label{eq:Born_prob}
\end{equation}
For each POVM, the associated classical Fisher-information matrix $F(\boldsymbol{\theta})$ is given by Eq.~\eqref{eq:classical_FIM_def}.
Optimizing over all POVMs yields the quantum Fisher-information matrix (QFIM) $Q(\boldsymbol{\theta})$, whose elements can be expressed in terms of the symmetric logarithmic derivatives (SLDs) $\hat{L}_a$ defined implicitly by
\begin{equation}
    \frac{\partial \hat{\rho}(\boldsymbol{\theta})}{\partial \theta_a}
    = \frac{1}{2}\big( \hat{\rho}(\boldsymbol{\theta}) \hat{L}_a
                      + \hat{L}_a \hat{\rho}(\boldsymbol{\theta}) \big).
    \label{eq:SLD_def}
\end{equation}
The QFIM elements are then
\begin{equation}
    Q_{ab}(\boldsymbol{\theta})
    = \frac{1}{2}\,\mathrm{Tr}\big[
        \hat{\rho}(\boldsymbol{\theta})
        \big( \hat{L}_a \hat{L}_b + \hat{L}_b \hat{L}_a \big)
      \big].
    \label{eq:QFIM_def}
\end{equation}
The quantum Cram\'{e}r--Rao bound (QCRB) states that the covariance matrix of any unbiased estimator satisfies
\begin{equation}
    \mathrm{Cov}(\hat{\boldsymbol{\theta}})
    \;\succeq\; \frac{1}{\mu}\,Q(\boldsymbol{\theta})^{-1},
    \label{eq:QCRB}
\end{equation}
for any physically allowed measurement strategy.
Since $F(\boldsymbol{\theta}) \le Q(\boldsymbol{\theta})$ in the matrix sense, the QFIM sets the ultimate precision limits allowed by quantum mechanics.

In the single-parameter case $(p=1)$, the QCRB is always saturable in the asymptotic limit $\mu \to \infty$.
For multiple parameters, however, the situation is more subtle: in general, there may not exist a single POVM that simultaneously achieves the quantum-limited precision for all components of $\boldsymbol{\theta}$. A necessary condition for the simultaneous saturation of the QCRB for all parameters is the weak commutativity condition \(\mathrm{Tr}(\rho[\hat{L}_a,\hat{L}_b]) = 0\). For the existence of a single-copy measurement achieving \(F(\boldsymbol{\theta}) = Q(\boldsymbol{\theta})\), a sufficient condition is the strong commutativity \([\hat{L}_a, \hat{L}_b] = 0\). However, under the weak commutativity condition, the QCRB can be asymptotically saturated for all parameters simultaneously using collective measurements on infinitely many copies.

\subsection{Figures of merit and local identifiability for joint estimation}

In a multiparameter scenario, there is no unique scalar figure of merit, since the covariance matrix encodes trade-offs between the variances and covariances of different parameters.  A common operational choice is a weighted sum of variances,
\begin{equation}
    \Delta^2_W(\hat{\boldsymbol{\theta}})
    = \mathrm{Tr}\!\left[W\,\mathrm{Cov}(\hat{\boldsymbol{\theta}})\right],
\end{equation}
where $W\succeq0$ is a user-chosen cost matrix.  For a specified measurement and a locally identifiable statistical model, the classical CRB gives
\begin{equation}
    \Delta^2_W(\hat{\boldsymbol{\theta}})
    \ge \frac{1}{\mu}\,\mathrm{Tr}\!\left[W\,F(\boldsymbol{\theta})^{-1}\right].
    \label{eq:weighted_crb_operational}
\end{equation}
This quantity has a direct operational meaning only when $F$ is full rank on the parameter subspace of interest.

This rank condition is especially important for the two-outcome photon-counting model used below.  If one measures a single binary probability $P_0(\boldsymbol{\theta})$ and its complement $1-P_0(\boldsymbol{\theta})$, then
\begin{equation}
    F_{ab}^{(1)}(\boldsymbol{\theta})=
    \frac{\partial_a P_0\,\partial_b P_0}{P_0(1-P_0)},
    \label{eq:rank_one_binary_fisher}
\end{equation}
which is an outer product and therefore satisfies
\begin{equation}
    \det F^{(1)}(\boldsymbol{\theta})=0
\end{equation}
for two parameters at a fixed operating point.  Thus, a single binary setting cannot by itself support local simultaneous estimation of $(\beta,x)$; it can estimate one parameter with the other calibrated, or it can be used as one component of a multi-setting experiment. 
Large values of the Fisher-information elements should not be confused with the ability to perform genuine simultaneous multiparameter estimation. The single-setting Fisher-information maps considered here quantify the local sensitivity of the measurement outcomes to parameter variations. They therefore provide useful sensitivity diagnostics, but they do not by themselves constitute an operational multiparameter estimation protocol because the associated Fisher matrix has rank one and the underlying statistical model is not locally identifiable with respect to both parameters. Accordingly, genuine local multiparameter estimation requires a full-rank Fisher matrix, which in the present framework is obtained only after combining multiple controlled reference settings.

For simultaneous estimation we therefore use a total Fisher matrix
\begin{equation}
    F_{\rm tot}(\boldsymbol{\theta})=
    \sum_{j=1}^{m}\nu_j F^{(j)}(\boldsymbol{\theta}),
    \qquad \sum_{j=1}^{m}\nu_j=1,
    \label{eq:total_fisher_settings}
\end{equation}
where $j$ labels controlled settings, such as different reference phases, interrogation times, or circuit configurations, and $\nu_j$ is the fraction of shots assigned to setting $j$.  
While the use of multiple reference phases is formally
related to standard phase-multiplexing strategies, the
purpose of the additional settings in the present work is
different. Here they are introduced primarily to restore
local identifiability of the joint $(\beta,x)$ estimation
problem. In other words, the additional settings are not
used merely to improve phase sensitivity, but rather to
generate non-collinear score vectors and thereby produce
a full-rank Fisher-information matrix.

A meaningful two-parameter CRB then requires $\det F_{\rm tot}>0$.  In the absence of a problem-specific cost, we use dimensionless parameters $(\beta\hbar\omega_0,x)$ and the simple choice $W=\mathbb{I}$; the plotted single-setting Fisher elements are interpreted as information landscapes, whereas the scalar joint-estimation cost is $\mathrm{Tr}[W F_{\rm tot}^{-1}]$.

\section{Physical model and analytical Fisher information}
\label{sec:model}

\subsection{Mach--Zehnder interferometer with a thermal atom}

\begin{figure}[t]
    \centering
    \includegraphics[width=0.85\columnwidth]{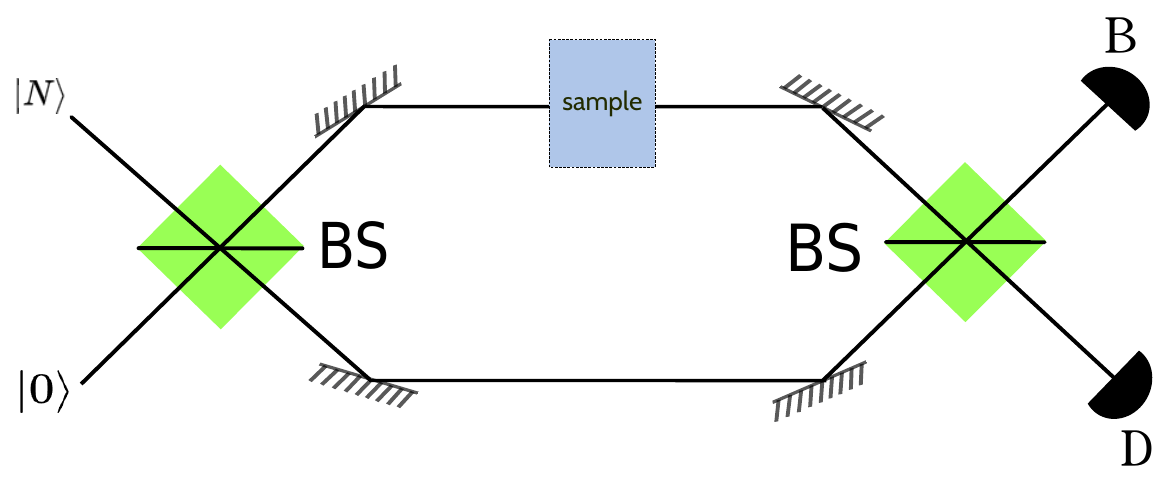}
    \caption{
        Schematic of the Mach--Zehnder interferometer (MZI) built with nonlinear beam splitters (BS) used in this work.
        A NOON state of $N$ photons enters the interferometer and, in one of the arms,
        the light field interacts dispersively with a two-level atom (sample) prepared
        in a thermal state at inverse temperature $\beta$. $D$ and $B$ denote detector labels corresponding to the two logical photon-counting outcomes.
    }
    \label{fig:setup}
\end{figure}

We consider a Mach--Zehnder interferometer (MZI) in which a light field interacts dispersively with a two-level atom located in one of the interferometer arms, as schematically illustrated in Fig.~\ref{fig:setup}.
The atomic transition $\ket{g}\leftrightarrow\ket{e}$ has energy splitting $\hbar\omega_0$. The dispersive light--matter coupling is described by the effective Hamiltonian in Eq.~\eqref{eq:dispersive_H}, and the corresponding unitary operator is given by Eq.~\eqref{eq:dispersive_U}. This interaction imprints a photon-number-dependent phase on the optical field only when the atom is in the excited state.

The atom is initially prepared in a thermal (Gibbs) state characterized by the inverse temperature
$\beta = 1/(k_{\mathrm{B}}T)$.  We keep $\beta$ in analytical derivatives to display the factor $\hbar\omega_0$ explicitly, while all plotted temperature axes use the dimensionless coordinate $b=\beta\hbar\omega_0$ (equivalently $b=\beta$ in units $\hbar\omega_0=1$).  The state is
\begin{equation}
    \hat{\rho}_{\mathrm{A}}(\beta)
    = \frac{\ket{g}\!\bra{g}+e^{-\beta\hbar\omega_0}\ket{e}\!\bra{e}}{Z_\beta},
    \qquad
    Z_\beta = 1 + e^{-\beta\hbar\omega_0}.
    \label{eq:rho_atom}
\end{equation}
Equivalently, in the ordered basis $\{\ket{g},\ket{e}\}$ this state is diagonal with entries
$p_g=1/Z_\beta$ and $p_e=e^{-\beta\hbar\omega_0}/Z_\beta$.
The optical input of the MZI is taken to be a NOON state of $N$ photons,
\begin{equation}
    \ket{\psi_{\mathrm{in}}}
    = \frac{1}{\sqrt{2}}\big(\ket{N,0}+\ket{0,N}\big),
    \label{eq:NOON}
\end{equation}
which provides the maximal phase sensitivity allowed by the optical resources.
Following Ref.~\cite{akamatsu2025fundamental}, we prepare the NOON state and
consider the action of a nonlinear beam splitter on this state, which leads
to the following transformations:
\begin{equation}
\begin{aligned}
    \ket{N}_a \otimes \ket{0}_b \equiv \ket{N,0}
        &\;\longrightarrow\;
        \frac{1}{\sqrt{2}}\big(\ket{N,0} + \ket{0,N}\big),\\[2mm]
    \ket{0}_a \otimes \ket{N}_b \equiv \ket{0,N}
        &\;\longrightarrow\;
        \frac{1}{\sqrt{2}}\big(\ket{0,N} - \ket{N,0}\big).
\end{aligned}
\label{eq:BS-nonlinear}
\end{equation}

The factor of two in the interference phase is a convention that must be fixed before comparing the analytical MZI and the circuit.  We use $x$ as the half-phase coordinate: the physical phase acquired by one photon in the excited atomic branch is $\varphi=\chi t=2x$, so an $N$-photon NOON component acquires the relative phase $2Nx$.  With the inverse nonlinear beam splitter used for recombination,
\begin{multline}
	\frac{e^{-i2Nx}|N,0\rangle+|0,N\rangle}{\sqrt{2}} \\
	\longmapsto e^{-iNx}\Big[\cos(Nx)|N,0\rangle-i\sin(Nx)|0,N\rangle\Big],
	\label{eq:second_bs_phase_convention}
\end{multline}
for the excited atomic branch, whereas the ground branch remains $|N,0\rangle$ at the selected output port.

After tracing out the atomic degrees of freedom and one of the two interferometer arms, the reduced state becomes

\begin{equation}
\begin{aligned}
\rho_{D}
&=
\sum_{n=0}^{1}
p_n
\Bigl[
\cos^{2}\!\left(nNx\right)
|0\rangle_a\langle0|
+
\sin^{2}\!\left(nNx\right)
|N\rangle_a\langle N|
\Bigr].
\end{aligned}
\label{new}
\end{equation}

This expression makes explicit that, for the nonlinear beam splitter considered in our protocol, the output state has support only on the two-dimensional subspace spanned by $|0\rangle$ and $|N\rangle$. Consequently, although photon counting is, in general, a multi-outcome measurement, only the outcomes corresponding to $0$ and $N$ detected photons occur with nonzero probability in the present scheme. The effective binary measurement therefore follows directly from the interferometer dynamics and does not arise from any post-selection or artificial restriction of the measurement outcomes.

For resource accounting, the protocol requires preparation of the NOON or NOON-encoded input, implementation of the two nonlinear or compiled beam-splitter operations, preparation of the thermal atomic state (with an ancillary qubit in the circuit realization), one unmeasured output-arm degree of freedom, and photon-number-resolving detection or its logically equivalent binary readout. For the multi-setting experiment, the total shot budget is distributed among the controlled settings according to the fractions $\nu_j$ in Eq.~\eqref{eq:Ftot_model}.
The probability of obtaining the detector event $D_0$ is therefore
$P_0^{(D)}=p_g+p_e\cos^2(Nx)$.
All Fisher-information formulas below use the phase convention $2x=\chi t$ fixed in Eq.~\eqref{eq:second_bs_phase_convention}.

\subsection{Output probabilities and thermal visibility}

After the first (nonlinear) beam splitter, the dispersive interaction acts only on the upper arm of the interferometer.
In our implementation, the light--matter coupling is governed by
$\hat H=\hbar\chi\,\hat a^\dagger\hat a\,\sigma_+\sigma_-=\hbar\chi\,\hat a^\dagger\hat a\,|e\rangle\langle e|$,
so that a phase is imprinted \emph{only} when the atom is excited.
After recombination at the second (also nonlinear) beam splitter, we perform photon-number measurements at the output ports. 
We denote by $D_0$ the detector event corresponding to
the output state $|N,0\rangle_{\rm out}$ and by $D_N$
the detector event corresponding to
$|0,N\rangle_{\rm out}$. Here the subscript ``0''
labels the detector event and should not be confused
with the vacuum state.

Tracing over the atomic degree of freedom yields the classical probabilities
\begin{equation}
\begin{aligned}
  P_0(\beta,x)
  &= \frac{1+\mathcal{V}(\beta)\cos(2Nx)}{1+\mathcal{V}(\beta)},\\
  P_N(\beta,x)
  &= 1-P_0(\beta,x)
   = \frac{2\mathcal{V}(\beta)}{1+\mathcal{V}(\beta)}\sin^2(Nx),
\end{aligned}
\label{eq:probabilities_Vstd}
\end{equation}
where we have defined the \emph{fringe visibility}
\begin{equation}
  \mathcal{V}(\beta)\equiv \mathcal{V}_{\rm std}
  =\frac{P_{\max}-P_{\min}}{P_{\max}+P_{\min}}
  =\frac{e^{-\beta\hbar\omega_0}}{2+e^{-\beta\hbar\omega_0}},
  \label{eq:visibility}
\end{equation}
with $P_{\max}=1$ (at $Nx=k\pi$) and
$P_{\min}=\frac{1-\mathcal{V}(\beta)}{1+\mathcal{V}(\beta)}$ (at $Nx=(k+\tfrac12)\pi$),
for any integer $k$.
Note that $\mathcal{V}(\beta)$ modulates the fringe contrast and is set by the thermal mixing between a
phase-imprinting branch ($|e\rangle$) and a phase-free branch ($|g\rangle$).
This visibility is a monotonic function of $\beta$: it approaches $0$ in the cold (non-inverted) limit
$\beta\to+\infty$ (where $p_e\to0$ and $P_0\to1$), takes the intermediate value $\mathcal{V}(0)=1/3$ at infinite
temperature, and approaches $1$ in the cold inverted limit $\beta\to-\infty$ (where $p_e\to1$ and
$P_0\to\frac12[1+\cos(2Nx)]=\cos^2(Nx)$).
Therefore, the fringe contrast encodes the temperature, including the sign of $\beta$ (e.g., $\mathcal{V}>1/3$
corresponds to $\beta<0$ and $\mathcal{V}<1/3$ to $\beta>0$).

\subsection{Analytical Fisher-information matrix and multi-setting completion}

Because the measurement outcomes $\{0,N\}$ correspond to orthogonal photon-counting projectors, the information analyzed in this section is the classical Fisher information of that specified measurement.  To make the identifiability structure explicit, we allow a controlled reference phase $\phi_j$ in setting $j$, so that
\begin{equation}
\begin{aligned}
  P_0^{(j)}(\beta,x)
  &= \frac{1+\mathcal{V}(\beta)\cos(2Nx+\phi_j)}{1+\mathcal{V}(\beta)},\\
  P_N^{(j)}(\beta,x)&=1-P_0^{(j)}(\beta,x).
\end{aligned}
\label{eq:probabilities_setting_j}
\end{equation}
The zero-reference case $\phi_j=0$ reduces to Eq.~\eqref{eq:probabilities_Vstd}.  From Eqs.~\eqref{eq:probabilities_setting_j} and \eqref{eq:visibility}, the derivatives are
\begin{equation}
\begin{aligned}
  \partial_\beta P_0^{(j)}
  &= \frac{\mathcal{V}'(\beta)\,[\cos(2Nx+\phi_j)-1]}
          {[1+\mathcal{V}(\beta)]^2},\\
  \partial_x P_0^{(j)}
  &= -\frac{2N\,\mathcal{V}(\beta)}{1+\mathcal{V}(\beta)}\,\sin(2Nx+\phi_j),
\end{aligned}
\label{eq:partials_setting_j}
\end{equation}
with $\mathcal{V}'(\beta)=-\hbar\omega_0\mathcal{V}(\beta)[1-\mathcal{V}(\beta)]$.

The Fisher matrix for one setting is
\begin{equation}
  F_{ab}^{(j)}(\beta,x)
  = \frac{\partial_a P_0^{(j)}\,\partial_b P_0^{(j)}}
         {P_0^{(j)}[1-P_0^{(j)}]},
  \qquad \boldsymbol{\theta}=(\beta,x)^\top.
  \label{eq:FIM_def_setting_j}
\end{equation}
For $\phi_j=0$ this gives
\begin{align}
  F_{\beta\beta}^{(0)}
    &= \frac{2(\hbar\omega_0)^2\,\mathcal{V}(\beta)[1-\mathcal{V}(\beta)]^2\,\sin^2(Nx)}
            {[1+\mathcal{V}(\beta)]^2[1+\mathcal{V}(\beta)\cos(2Nx)]},
  \label{eq:Fbb}\\[4pt]
  F_{xx}^{(0)}
    &= \frac{8N^2\,\mathcal{V}(\beta)\,\cos^2(Nx)}
            {1+\mathcal{V}(\beta)\cos(2Nx)},
  \label{eq:Fxx}\\[4pt]
  F_{\beta x}^{(0)}
    &= -\frac{2N\hbar\omega_0\,\mathcal{V}(\beta)[1-\mathcal{V}(\beta)]\,\sin(2Nx)}
            {[1+\mathcal{V}(\beta)][1+\mathcal{V}(\beta)\cos(2Nx)]}.
  \label{eq:Fbx}
\end{align}

These are the diagnostic Fisher elements plotted below.  Since Eq.~\eqref{eq:FIM_def_setting_j} is an outer product, however, $\det F^{(j)}=0$ for every single binary setting. 

Because the photon-counting records obtained from different controlled settings are statistically independent, the overall likelihood factorizes and the corresponding Fisher information is additive. The total Fisher matrix is therefore obtained as the weighted sum of the Fisher matrices associated with the individual settings. Equivalently, the complete protocol can be represented as one enlarged-outcome measurement with outcome $(j,d)$, where $j$ is the known control setting and $d\in\{0,N\}$ is the detector result; the Fisher information of this joint distribution is precisely the weighted sum below.

A locally invertible two-parameter experiment must combine at least two non-collinear settings.  If $\mu_{\rm tot}$ shots are distributed with fractions $\nu_j$ ($\sum_j\nu_j=1$), the count-level information entering the asymptotic covariance of an estimator is
\begin{equation}
    F_{\rm tot}(\beta,x)=\mu_{\rm tot}\sum_j \nu_j F^{(j)}(\beta,x),
    \label{eq:Ftot_model}
\end{equation}
where $\det F_{\rm tot}>0$. The quantity $\sum_j \nu_j F^{(j)}$ is the Fisher matrix per experimental shot, whereas the total Fisher information scales linearly with the number of shots $\mu_{\rm tot}$, so that the corresponding Cram\'{e}r--Rao bound scales as $1/\mu_{\rm tot}$. 
Reference phases separated by quadrature, for example $\phi_j\in\{0,\pi/2,\pi\}$, make the score vectors for $(\beta,x)$ non-parallel over broad regions of the parameter plane. The choice of reference phases is not unique, and the set used here is a convenient, symmetric proof-of-principle choice guided by the periodicity of the interference fringe. Its purpose is to establish local identifiability by generating non-collinear score vectors, rather than to claim a unique or globally optimal experimental design. A separate optimization would require specifying an additional scalar objective---for example $\det F_{\rm tot}$, the minimum eigenvalue, the condition number, or a weighted value of $\mathrm{Tr}[F_{\rm tot}^{-1}]$---together with a prior parameter region and a shot-allocation constraint. Because different choices define different design problems, such optimization is not required for the identifiability result established here. In this sense, the controlled phases enrich the measurement record and overcome the rank deficiency inherent to any single binary photon-counting setting.

The operational scalar quantity used for joint estimation is therefore the weighted CRB cost
\begin{equation}
    \mathcal{C}_W(\beta,x)=\mathrm{Tr}\!\left[W F_{\rm tot}(\beta,x)^{-1}\right],
    \label{eq:joint_cost_model}
\end{equation}
not the determinant of a single-setting Fisher matrix.

Given a count record $\{n_0^{(j)},n_N^{(j)}\}_{j=1}^K$, the corresponding likelihood is
\begin{equation}
    \mathcal{L}(\beta,x)=
    \prod_{j=1}^{K}
    \left[P_0^{(j)}(\beta,x)\right]^{n_0^{(j)}}
    \left[1-P_0^{(j)}(\beta,x)\right]^{n_N^{(j)}}.
    \label{eq:multisetting_likelihood}
\end{equation}
The maximum-likelihood estimator is
\begin{equation}
    (\hat\beta,\hat x)_{\rm MLE}
    =\arg\max_{\beta,x}\log\mathcal{L}(\beta,x).
    \label{eq:mle_estimator}
\end{equation}
Under the usual regularity conditions and within a locally identifiable region, the covariance of this estimator approaches $F_{\rm tot}^{-1}$ in the large-sample limit.  This establishes that the multi-setting Fisher matrix is not merely a visualization tool; it is the local curvature of an explicit estimator built from the photon-counting record.

The expressions above make explicit how temperature enters the photon-counting statistics through the monotonic visibility $\mathcal{V}(\beta)$.  Information about $x$ is suppressed when $\mathcal{V}\to0$, because the phase-imprinting branch is unoccupied, whereas thermometric information is suppressed when $\mathcal{V}'(\beta)\to0$, which occurs in both nearly pure limits.  Thus the same likelihood can be phase-dominated in the inverted regime and temperature-sensitive only in an intermediate window where the visibility slope is appreciable.

\subsection{Extension to negative temperatures}
\label{subsec:negative_T}

The analytical expressions derived above remain valid for negative inverse temperatures ($\beta<0$), corresponding to population inversion in systems with bounded spectra \cite{braun2013negative,frenkel2015gibbs,cerino2015consistent,Assis2019,nettersheim2022power}. Such states have attracted considerable interest in quantum thermodynamics because they may serve as resources for enhanced work extraction and quantum metrology \cite{ivanchenko2015quantum,Assis2019,nettersheim2022power,xi2017quantum}.

For the present interferometric scheme, the visibility
\begin{equation}
  \mathcal{V}(\beta)=
  \frac{e^{-\beta\hbar\omega_0}}
       {2+e^{-\beta\hbar\omega_0}},
\end{equation}
remains strictly positive and therefore negative temperatures do not induce a $\pi$ phase shift of the interference fringes. Instead, population inversion increases the weight of the phase-imprinting branch $|e\rangle$, enhancing the fringe contrast. Consequently, the output probabilities
\begin{equation}
  P_0(\beta,x)=
  \frac{1+\mathcal{V}(\beta)\cos(2Nx)}
       {1+\mathcal{V}(\beta)}
\end{equation}
retain the same phase dependence while their modulation amplitude increases as $\beta$ becomes more negative.

This behavior is directly reflected in the Fisher-information matrix. Since
$\mathcal{V}'(\beta)
=
-\hbar\omega_0\,\mathcal{V}(\beta)
[1-\mathcal{V}(\beta)]$,
the phase sensitivity increases monotonically with
$\mathcal{V}(\beta)$ and reaches its maximum in the cold inverted limit
($\beta\rightarrow-\infty$), where the interferometer recovers the ideal NOON-state fringes. By contrast, the thermometric sensitivity is governed by
$\mathcal{V}'(\beta)$ and is therefore maximal at intermediate temperatures, vanishing in both limits
$\beta\rightarrow\pm\infty$.

Thus, in the present protocol, negative temperatures do not modify the fringe phase but enhance the interferometric contrast through population inversion. The visibility $\mathcal{V}(\beta)$ therefore plays the role of the operational resource controlling the information available for the joint estimation of $(\beta,x)$.

\subsection{Quantum Fisher information and measurement optimality}

The Fisher-information analysis presented in this work is measurement specific because it is based on the binary photon-counting statistics $\{D_0,D_N\}$. We therefore compare the corresponding classical Fisher matrix with the QFIM of the reduced detector state $\rho_D$ in Eq.~\eqref{new}. This comparison characterizes optimality within the effective binary-output model; it is not a claim about the unreduced atom--field output state.

Because $\rho_D$ remains diagonal in a parameter-independent eigenbasis, its QFIM reduces to

\begin{equation}
Q_{ab}
=
\sum_{\substack{i,j=\{0,N\}\\ P_i+P_j\neq 0}}
2\,
\frac{
\mathrm{Re}
\!\left(
\langle i|\partial_a\rho_D|j\rangle
\langle j|\partial_b\rho_D|i\rangle
\right)
}
{P_i+P_j},
\label{eq:QFI_general}
\end{equation}
which coincides exactly with the classical Fisher matrix
associated with the photon-counting probabilities,

\begin{equation}
Q_{ab}=F_{ab}.
\end{equation}

Therefore, the binary photon-counting POVM saturates the QFIM of the reduced detector state for each setting. Moreover, because the eigenbasis is parameter independent, the symmetric logarithmic derivatives are diagonal in the same basis and commute on the support of $\rho_D$. The compatibility condition is therefore satisfied within this reduced operational model. This statement does not assert global optimality for the unreduced atom--field state. Within the effective output model, the limitation discussed below does not originate from measurement suboptimality.

For a single binary photon-counting setting, the classical
Fisher matrix is

\begin{equation}
F^{(1)}_{ab}
=
\frac{
\partial_a P_0\,
\partial_b P_0
}
{
P_0(1-P_0)
},
\end{equation}

which is an outer product and therefore satisfies

\begin{equation}
\det F^{(1)}=0.
\end{equation}

As discussed in Sec.~II.C, $F^{(1)}$ has rank one and cannot support local simultaneous estimation of $(\beta,x)$. Within the reduced binary-output model, this rank deficiency is not caused by a suboptimal readout. Rather, it reflects the fact that a single binary setting provides only one independent score direction in the two-dimensional parameter space.

The multi-setting protocol introduced in this work
overcomes this limitation by combining several controlled
reference phases, which generate non-collinear score
vectors and produce a full-rank Fisher matrix. In this
sense, the controlled phases play a dual role: they enrich
the measurement record and lift the rank deficiency
inherent to any single binary photon-counting setting.

The role of the multi-setting protocol is therefore not to increase the QFIM of the reduced detector state for a fixed setting. Instead, it combines linearly independent score directions generated by different controlled reference phases, thereby restoring local identifiability through a full-rank Fisher matrix.

\section{Quantum-circuit implementation}
\label{sec:circuit}

In this section we show how the interferometric model of Sec.~\ref{sec:model} can be mapped onto a quantum circuit acting on qubits.
This provides a digital platform to emulate the photon-counting likelihood and to benchmark the analytical Fisher-information expressions derived previously.
It also makes clear how the two logical output-port outcomes are implemented as standard projective measurements in the computational basis. The resulting measurement is the specified photon-counting measurement used throughout the paper. Its equality with the QFIM established in Sec.~\ref{sec:model} applies to the reduced detector state and should not be interpreted as a proof of global optimality for the unreduced atom--field output state.

\subsection{Encoding of optical modes and atom}

The relevant Hilbert space of the interferometer can be restricted to the two-dimensional subspace spanned by the NOON components $\ket{N,0}$ and $\ket{0,N}$.
We identify this two-dimensional subspace with one \emph{logical light qubit} having basis
\begin{equation}
    \ket{0}_\mathrm{L} \equiv \ket{N,0},\qquad
    \ket{1}_\mathrm{L} \equiv \ket{0,N}.
\end{equation}

The logical encoding introduced above is independent of the physical implementation. In the four-qubit circuit of Fig.~\ref{fig:MZ_circuit}, which realizes the case $N=1$, this single logical qubit is represented physically by the dual-rail subspace of $q_2q_3$, with $\ket{0}_{\rm L}=\ket{10}_{q_2q_3}$ and $\ket{1}_{\rm L}=\ket{01}_{q_2q_3}$, up to the path-label convention used by the compiled circuit. For larger photon numbers, each branch would require a larger register.

In this encoding, the NOON state \eqref{eq:NOON} becomes simply
\begin{equation}
    \ket{\psi_{\mathrm{in}}}
    = \frac{1}{\sqrt{2}}\big(\ket{0}_\mathrm{L} + \ket{1}_\mathrm{L}\big),
\end{equation}
which can be prepared from $\ket{0}_\mathrm{L}$ by a single Hadamard gate
$\hat{H}$ on the light qubit.

The two-level atom is represented by a second qubit, the \emph{atomic qubit}, with
\begin{equation}
    \ket{g} \equiv \ket{0}_\mathrm{A},\qquad
    \ket{e} \equiv \ket{1}_\mathrm{A},
\end{equation}
and Hamiltonian splitting $\hbar\omega_0$.
The dispersive interaction \eqref{eq:dispersive_H} corresponds, in this effective two-qubit picture, to a controlled phase rotation on the light qubit conditioned on the atomic excitation.
With the convention $2x=\chi t$, the excited branch receives the relative phase $2Nx$.  Up to a global phase, this can be implemented as a controlled-$R_z$ gate
\begin{equation}
    \hat{U}_\mathrm{disp}(x)
    = \ket{g}\!\bra{g} \otimes \mathbb{I}_\mathrm{L}
    + \ket{e}\!\bra{e} \otimes \hat{R}_z(2Nx),
    \label{eq:U_disp_circuit}
\end{equation}
where
\begin{equation}
    \hat{R}_z(\phi)
    = e^{-i\phi \hat{\sigma}_z/2}
\end{equation}
acts on the encoded path qubit.  In the ideal logical two-level description, the MZI is therefore
\begin{equation}
    \hat{U}_\mathrm{MZI}(x)
    = B_2\,\hat{U}_\mathrm{disp}(x)\,B_1,
    \label{eq:UMZI_circuit}
\end{equation}
where $B_1$ prepares the NOON-path superposition and $B_2=B_1^\dagger$ recombines it according to Eq.~\eqref{eq:second_bs_phase_convention}.  In the four-qubit circuit below these logical beam splitters are not single Hadamards on an isolated qubit; they are compiled as Hadamard--CNOT and CNOT--Hadamard--CNOT blocks acting on the two encoded path qubits.

\subsection{Preparation of the thermal atomic state}

The initial atomic state is the thermal state \eqref{eq:rho_atom}, which is diagonal in the $\{\ket{g},\ket{e}\}$ basis.
In a quantum-circuit framework, the same reduced thermal state can be realized in two operationally equivalent ways:

\begin{enumerate}
    \item \emph{Classical sampling of initial states.}
    For each run of the experiment, the atomic qubit is initialized either in $\ket{g}$ or in $\ket{e}$ with probabilities
    \begin{equation}
        p_g(\beta)
        = \frac{1}{Z_\beta},\qquad
        p_e(\beta)
        = \frac{e^{-\beta\hbar\omega_0}}{Z_\beta},
    \end{equation}
    where $Z_\beta$ is given in Eq.~\eqref{eq:rho_atom}.
    The statistics are then obtained by averaging the measurement outcomes over many shots, weighted by these classical probabilities.

    \item \emph{Coherent preparation with an ancilla.}
    Alternatively, one may entangle the atomic qubit with an auxiliary qubit and then trace out the ancilla.
    For instance, preparing an entangled state
    \(
    \sqrt{p_g}\,\ket{0}_\mathrm{anc}\ket{g}
    + \sqrt{p_e}\,\ket{1}_\mathrm{anc}\ket{e}
    \)
    and discarding the ancilla yields the desired mixed state \eqref{eq:rho_atom} on the atom.
\end{enumerate}

In our numerical simulations we adopt the second strategy.

\subsection{Quantum circuit for the Mach--Zehnder scheme}
\label{subsec:circuit_fig}

The interferometric model of Sec.~\ref{sec:model} can be implemented on a
qubit-based quantum processor using the four-qubit circuit shown in
Fig.~\ref{fig:MZ_circuit}.
All qubits are initialized in the state $\ket{0}$.
Qubit $q_1$ encodes the two-level atom, while $q_2$ represents the
interferometer output mode that is finally measured.
Qubit $q_3$ plays the role of the reference arm of the interferometer, and
$q_0$ is an ancilla used to prepare the thermal state of the atom.

\paragraph*{Thermal state preparation.}
The atomic thermal state $\hat{\rho}_{\mathrm{A}}(\beta)$ in Eq.~\eqref{eq:rho_atom}
is generated by first applying a single-qubit rotation $\hat{R}(\beta)$ to the
ancilla $q_0$,
\begin{equation}
\begin{aligned}
    \hat{R}(\beta)
    &=
    \begin{pmatrix}
        \sqrt{p_g(\beta)} & -\sqrt{p_e(\beta)} \\
        \sqrt{p_e(\beta)} & \sqrt{p_g(\beta)}
    \end{pmatrix},\\
    p_g(\beta)&=\frac{1}{1+e^{-\beta\hbar\omega_0}},
    \qquad
    p_e(\beta)=\frac{e^{-\beta\hbar\omega_0}}{1+e^{-\beta\hbar\omega_0}}.
\end{aligned}
\end{equation}
This prepares the superposition
$\hat{R}(\beta)\ket{0}_{q_0}=\sqrt{p_g}\ket{0}_{q_0}+\sqrt{p_e}\ket{1}_{q_0}$.
A subsequent CNOT gate with $q_0$ as control and $q_1$ as target entangles the
two qubits, encoding the classical mixture of ground and excited states in the
reduced state of $q_1$.
Tracing out (or simply ignoring) the ancilla then leaves the atomic qubit
$q_1$ in the desired thermal state $\hat{\rho}_{\mathrm{A}}(\beta)$.

\paragraph*{Preparation of the NOON-encoded input.}
The interferometer input state for $N=1$ is the NOON-like state
$(\ket{01}+\ket{10})/\sqrt{2}$ in the two path modes.  One convenient preparation is to start from $\ket{00}_{q_2q_3}$, apply $X$ to $q_3$, apply $H$ to $q_2$, and then apply a CNOT with $q_2$ as control and $q_3$ as target, which maps
$\ket{01}\mapsto\ket{01}$ and $\ket{11}\mapsto\ket{10}$ and therefore produces
$(\ket{01}+\ket{10})/\sqrt{2}$.  Equivalent compiled state-preparation blocks give the same logical input.  The subsequent interferometric beam splitters are represented by Hadamard-type rotations in the encoded path basis.

\paragraph*{Dispersive interaction and second beam splitter.}
The dispersive atom--field interaction of Eq.~\eqref{eq:dispersive_H} is realized as
a controlled phase gate between the atomic qubit $q_1$ and the light qubit
$q_2$.
In the circuit this is represented by a controlled-$U$ gate, where
\begin{equation}
    \hat{U}(x)
    =
    \begin{pmatrix}
        1 & 0 \\
        0 & e^{-i2x}
    \end{pmatrix},
\end{equation}
acts on $q_2$ whenever $q_1$ is in the excited state $\ket{1}$, ensuring that only
one arm of the interferometer acquires the physical dispersive phase $2x=\chi t$.
The second beam splitter of the Mach--Zehnder interferometer is then modeled
by a final CNOT gate with $q_2$ as control and $q_3$ as target, followed by a Hadamard gate on $q_2$ and another CNOT gate in the same configuration as the first one.

\paragraph*{Measurement and statistics.}
At the end of the circuit the light qubit $q_2$ is measured in the
computational basis, yielding a classical bit $c_0$ that records whether the
photons exit through the reference port ($\ket{0}_{q_2}$) or the complementary
port ($\ket{1}_{q_2}$). The final operations map the logical output-port information onto $q_2$. Consequently, although $q_2$ and $q_3$ jointly encode the logical state during the circuit evolution, the measured $q_2$ statistics are obtained after marginalizing over the unmeasured qubit $q_3$ and reproduce the effective binary photon-counting probabilities.
Repeating the circuit for many shots and for different pairs $(\beta,x)$
reconstructs the probabilities $P_0(\beta,x)$ and $P_1(\beta,x)$, which, as
discussed in Secs.~\ref{sec:model} and \ref{sec:results}, agree with the
analytical prediction of Eq.~\eqref{eq:probabilities_Vstd} and allow us to estimate
the entries of the Fisher-information matrix.
\begin{figure}[t]
    \centering
    \resizebox{1\linewidth}{!}{
    \begin{quantikz}
        \lvert q_0 \rangle & \gate{R}\gategroup[2, steps=2, style={dashed, rounded corners, fill=blue!10},
            background]{state preparation} & \ctrl{1} \\
        \lvert q_1 \rangle & \qw      & \targ{}  & \qw      & \ctrl{1}\gategroup[2, steps=1, style={dashed, rounded corners, fill=red!10}, background]{interaction} \\
        \lvert q_2 \rangle & \qw      & \gate{H}\gategroup[2, steps=2, style={dashed, rounded corners, fill=green!10}, background, label style={label position=below, yshift=-0.5cm}]{BS1} & \ctrl{1} & \gate{U} & \ctrl{1}\gategroup[2, steps=3, style={dashed, rounded corners, fill=green!10}, background, label style={label position=below, yshift=-0.5cm}]{BS2} & \gate{H} & \ctrl{1}  & \meter{}\\
        \lvert q_3 \rangle & \gate{X} & \qw      & \targ{1} & \qw      & \targ{}  & \qw      & \targ{}
    \end{quantikz}
    }
    \caption{%
        Quantum circuit used to simulate the Mach--Zehnder interferometer for
        $N=1$.
        Qubit $q_1$ encodes the thermal two-level atom, $q_2$ represents the
        interferometer output mode that is measured, and $q_3$ corresponds to
        the reference arm.
        Qubit $q_0$ is an ancilla used to prepare the atomic thermal state via
        the rotation $R$ and a CNOT.
        The dispersive interaction is implemented as a controlled-$U$ gate
        between $q_1$ and $q_2$, while Hadamard and CNOT gates model the beam splitters.
    }
    \label{fig:MZ_circuit}
\end{figure}

\subsection{Measurement, reconstruction, and estimator meaning}

The measurement stage consists of projective measurements in the computational basis of the light qubit.  The two computational outcomes are logical output-port labels: outcome $0$ corresponds to the logical state $\ket{0}_{\rm L}\equiv\ket{N,0}$ and outcome $N$ corresponds to $\ket{1}_{\rm L}\equiv\ket{0,N}$.  Thus the symbol ``0'' labels a detector port in the encoded two-dimensional subspace; it should not be confused with a physical zero-photon state.  With the atom unmeasured, the POVM is
\begin{equation}
    \hat{\Pi}_0 = \ket{0}_{\rm L}\!\bra{0}\otimes\mathbb{I}_{\rm A},
    \qquad
    \hat{\Pi}_N = \ket{1}_{\rm L}\!\bra{1}\otimes\mathbb{I}_{\rm A}.
\end{equation}
This POVM reproduces the probabilities in Eqs.~\eqref{eq:probabilities_Vstd} and \eqref{eq:probabilities_setting_j}.

We do not use the hardware data to claim global saturation of the quantum multiparameter Cram\'er--Rao bound.  For ideal simulations, the object reconstructed from counts is the classical Fisher matrix of the specified photon-counting likelihood.  For one binary setting it is rank one, as discussed in Sec.~\ref{sec:theory}; full two-parameter local estimation requires the multi-setting matrix $F_{\rm tot}$ in Eq.~\eqref{eq:Ftot_model}.  The single-setting entries $F_{\beta\beta}$, $F_{xx}$, and $F_{\beta x}$ remain useful diagnostics because they identify where the likelihood is most sensitive to temperature, phase, or correlated changes of both.

From the circuit, for each chosen pair $(\beta,x)$ and setting $j$, we obtain empirical frequencies
\begin{equation}
    \hat{P}_0^{(j)}(\beta,x)=\frac{n_0^{(j)}}{\mu_j},
    \qquad
    \hat{P}_N^{(j)}(\beta,x)=\frac{n_N^{(j)}}{\mu_j}.
\end{equation}
These frequencies converge to the theoretical probabilities in the large-shot limit.  We estimate the classical Fisher entries by using the analytical score functions in Eq.~\eqref{eq:partials_setting_j} together with the empirical probabilities.  This procedure tests the likelihood model and the impact of finite sampling and device noise; it is distinct from a full blind parameter-estimation experiment, which would require constructing estimators, computing empirical mean-square errors, and comparing those errors with $F_{\rm tot}^{-1}$ as a function of resources.

\subsection{Simulation details and comparison with analytical results}

The full protocol can be summarized as follows:
\begin{enumerate}
    \item Choose a grid of parameter values $(\beta,x)$ and, for joint estimation, a set of controlled settings $\{\phi_j\}$.
    \item For each grid point:
    \begin{enumerate}
        \item Prepare the thermal atomic state $\rho_{\mathrm{A}}(\beta)$ using the ancilla-assisted circuit described in Sec.~\ref{sec:circuit}, i.e., by applying the appropriate single-qubit rotation on the ancilla and a CNOT to entangle it with the atomic qubit, followed by tracing out (or discarding) the ancilla.
        \item Prepare the NOON-encoded state $(\ket{0}_\mathrm{L}+\ket{1}_\mathrm{L})/\sqrt{2}$ on the light qubit.
        \item Apply the interferometric unitary $\hat{U}_\mathrm{MZI}(x)$ in Eq.~\eqref{eq:UMZI_circuit}, optionally including the controlled reference phase $\phi_j$.
        \item Measure the light qubit in the computational basis and record the outcome.
        \item Repeat steps (a)--(d) a total of $\mu$ times to obtain the empirical probabilities $\hat{P}_0$ and $\hat{P}_N$.
    \end{enumerate}
    \item Use Eq.~\eqref{eq:FIM_def_setting_j} together with Eq.~\eqref{eq:partials_setting_j} to reconstruct the single-setting diagnostic elements and, when several settings are used, the full matrix $\hat F_{\rm tot}=\sum_j\nu_j\hat F^{(j)}$.
\end{enumerate}

In Sec.~\ref{sec:results} we compare the numerically reconstructed ideal-likelihood Fisher diagnostics $\hat{F}$ with the analytical expressions \eqref{eq:Fbb}--\eqref{eq:Fbx} derived from the visibility-based model.
We find excellent agreement across the parameter space, with deviations consistent with finite-sampling fluctuations.
This confirms that the quantum circuit faithfully reproduces the interferometric likelihood.  Full local joint estimation is obtained only after combining non-collinear settings into $F_{\rm tot}$.

\section{Results and discussion}
\label{sec:results}

The IBM Quantum implementation presented in this section is a hardware emulation of the photon-counting likelihood rather than a complete metrological validation. We use the hardware counts to reconstruct interference probabilities, obtain representative visibility-based point estimates, and assess device-induced distortions of the ideal-score Fisher diagnostics. The hardware run does not include blind joint estimation, an empirical estimator covariance matrix, or a hardware-level comparison with the full multi-setting Cram\'{e}r--Rao bound. The synthetic estimator benchmark in Sec.~\ref{subsec:multisetting_results} is therefore kept distinct from the IBM data.

\subsection{Benchmarking the interferometric statistics}

\begin{figure}[ht!]
    \centering
    \includegraphics[width=0.4\textwidth]{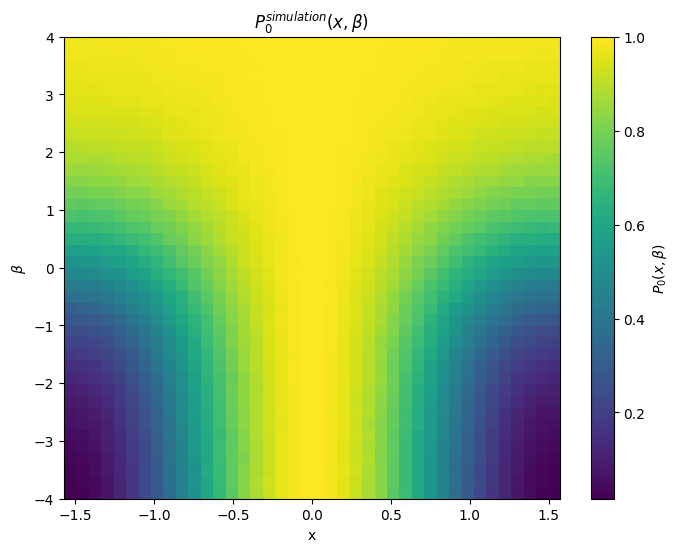}\hfill
    \includegraphics[width=0.4\textwidth]{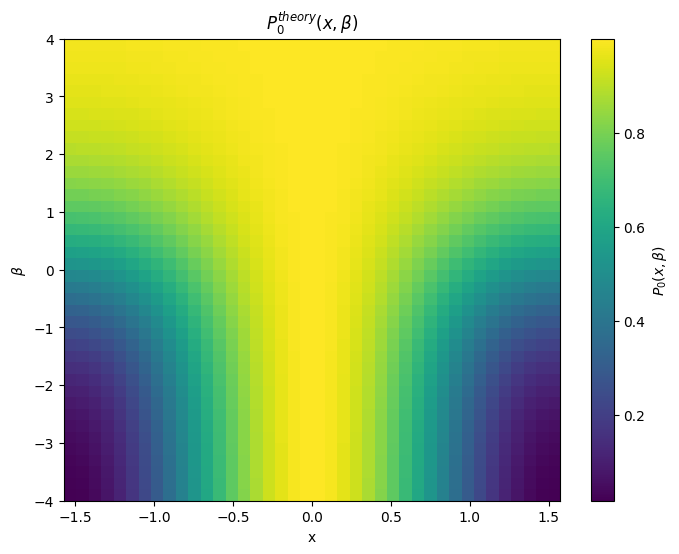}
    \caption{%
        Comparison between simulated and analytical interference probabilities.
        (a)~Simulation results $P_0(\beta,x)$ obtained from $10^4$ circuit shots.
        (b)~Analytical prediction from Eq.~\eqref{eq:probabilities_Vstd}.
        The color scale encodes the probability of detecting all photons at the reference output port.
    }
    \label{fig:P_comparison}
\end{figure}

We first compare the numerically simulated interference probabilities obtained from the quantum circuit with the analytical prediction of Eq.~\eqref{eq:probabilities_Vstd}.
Figure~\ref{fig:P_comparison} shows the three-dimensional surfaces of $P_0(\beta,x)$ from the circuit simulation [panel~(a)] and from the analytical model [panel~(b)].
The two plots are indistinguishable within the statistical fluctuations associated with the finite number of shots ($\mu=10^4$ in our runs).
As $\beta$ decreases (increasing inversion), the contrast increases: minima become deeper while the maxima remain fixed at $P_{\max}=1$.
The high degree of agreement between simulation and theory benchmarks the digital representation of the Mach--Zehnder interferometer introduced in Sec.~\ref{sec:circuit}.

\subsection{Information landscape from the analytical photon-counting FIM}

In all numerical landscapes, the temperature coordinate is the dimensionless variable $b=\beta\hbar\omega_0$; the labels $F_{\beta\beta}$ and $F_{\beta x}$ are retained to match the analytical derivatives, with $\hbar\omega_0=1$ in the plotted units.

\begin{figure}[ht!]
    \centering
    \includegraphics[width=0.48\textwidth]{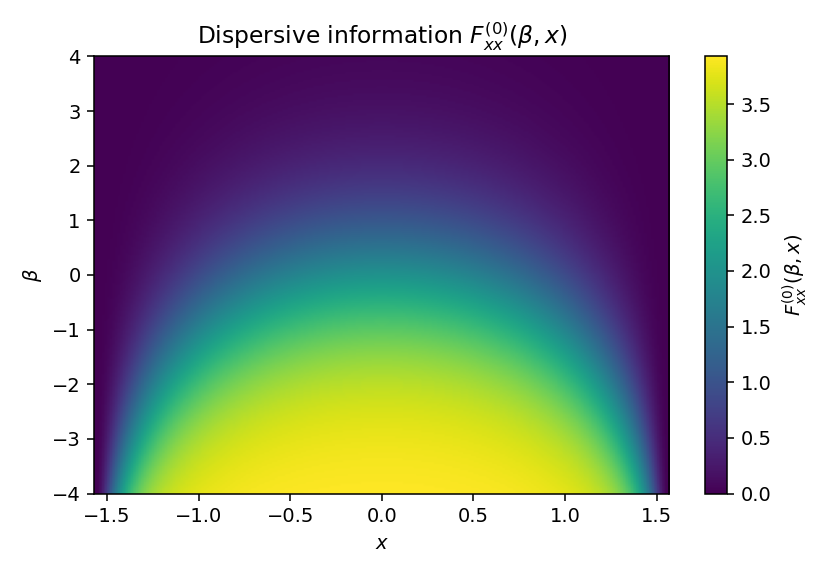}\hfill
    \includegraphics[width=0.48\textwidth]{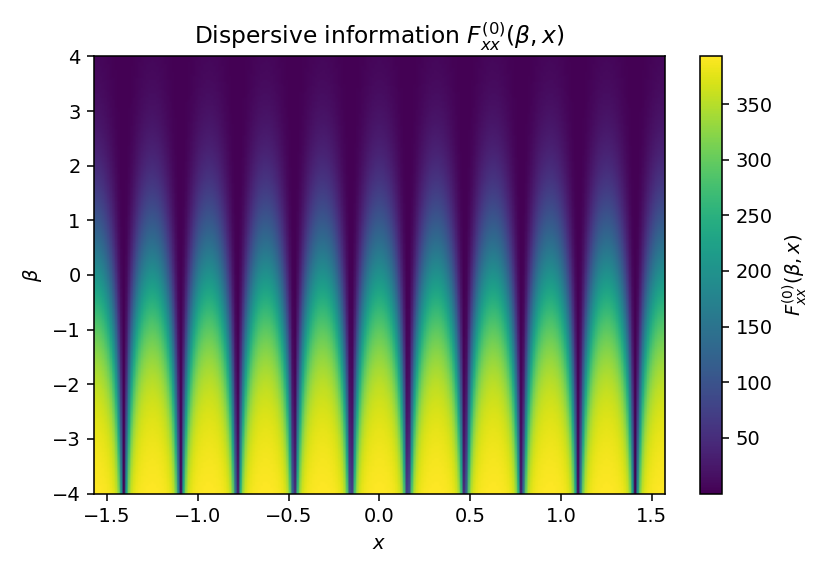}
    \caption{%
        Single-setting dispersive sensitivity diagnostic $F_{xx}^{(0)}(\beta,x)$ as a function of inverse temperature and interaction strength for NOON probes with
        (a)~$N=1$ and (b)~$N=10$ photons.
        The information is suppressed for large positive $\beta$ (cold non-inverted regime, $\mathcal{V}\to0$), and it approaches its maximum $F_{xx,\max}^{(0)}\simeq4N^2$ in the cold inverted limit $\beta\to-\infty$ where $\mathcal{V}\to1$.
        At $\beta=0$ one has $\mathcal{V}(0)=1/3$, so $F_{xx}^{(0)}$ remains finite.
        Increasing $N$ enhances the maximal information approximately as $N^2$ and produces narrower interference fringes along $x$, leading to multiple equivalent operating points.
    }
    \label{fig:Qxx_landscape}
\end{figure}

We now turn to the classical Fisher-information matrix of the photon-counting measurement.
The element $F_{xx}^{(0)}(\beta,x)$ is a single-setting sensitivity diagnostic for the dispersive phase $x$.  It becomes a single-parameter CRB only when $\beta$ is independently calibrated; when both parameters are unknown, the operational bound is obtained from the inverse of the multi-setting matrix $F_{\rm tot}$.
Figure~\ref{fig:Qxx_landscape} displays $F_{xx}^{(0)}$ as a function of $(\beta,x)$ for NOON probes with $N=1$ [panel~(a)] and $N=10$ [panel~(b)].
In both cases the phase information is \emph{suppressed} for large positive $\beta$ (cold non-inverted regime), because $\mathcal{V}(\beta)\to0$ and the atom remains in $|g\rangle$ so that no phase is imprinted.
As $\beta$ decreases and the excited population grows, $\mathcal{V}(\beta)$ increases and the phase information rises, reaching its maximal value $F_{xx,\max}^{(0)}\simeq 4N^2$ in the cold inverted limit $\beta\!\to\!-\infty$, where $\mathcal{V}(\beta)\to1$ and the interferometer recovers the ideal NOON fringes.
At $\beta\!=\!0$, the visibility is finite, $\mathcal{V}(0)=1/3$, and $F_{xx}^{(0)}$ remains finite (albeit reduced compared to the inverted regime).
Increasing $N$ enhances the overall scale of $F_{xx}^{(0)}$ by approximately a factor $N^2$ and produces narrower interference fringes along $x$, giving rise to several equivalent high-response points separated by $\pi/N$.

\begin{figure}[ht!]
    \centering
    \includegraphics[width=0.48\textwidth]{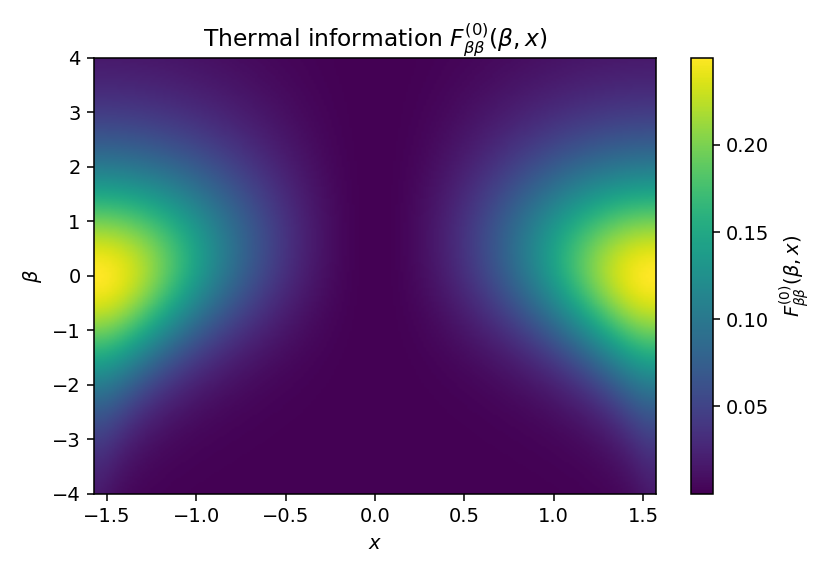}\hfill
    \includegraphics[width=0.48\textwidth]{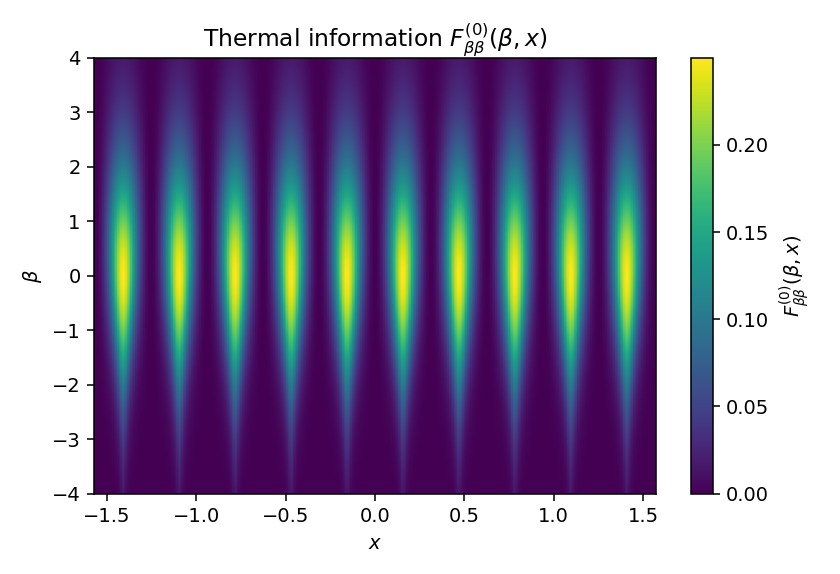}
    \caption{%
        Single-setting thermometric sensitivity diagnostic $F_{\beta\beta}^{(0)}(\beta,x)$ for NOON probes with
        (a)~$N=1$ and (b)~$N=10$ photons.
        The information is concentrated in an intermediate $\beta$ band where the visibility varies most rapidly (i.e., where $|\mathcal{V}'(\beta)|$ is largest), and it is suppressed for $|\beta|\to\infty$ where the atom becomes nearly pure.
        The overall scale of $F_{\beta\beta}^{(0)}$ is essentially independent of $N$, while larger $N$ produces a fine fringe structure along $x$ that mirrors the phase-sensitive lobes of $F_{xx}^{(0)}$.
    }
    \label{fig:Qbb_landscape}
\end{figure}

The complementary element $F_{\beta\beta}^{(0)}(\beta,x)$ is the corresponding single-setting sensitivity diagnostic for temperature (or, more precisely, for the inverse temperature $\beta$).
Its landscape is shown in Fig.~\ref{fig:Qbb_landscape} for $N=1$ [panel~(a)] and $N=10$ [panel~(b)].
In contrast with $F_{xx}^{(0)}$, the thermal information is concentrated in an intermediate band of $\beta$, where the visibility changes most rapidly.
Since $\mathcal{V}'(\beta)=-\hbar\omega_0\,\mathcal{V}(\beta)\big[1-\mathcal{V}(\beta)\big]$, the strongest thermometric response occurs when $\mathcal{V}(\beta)$ is neither too small nor too close to unity (roughly around $\mathcal{V}\sim 1/2$, i.e., $\beta\hbar\omega_0 \approx -\ln 2$).
For large $|\beta|$ the atom becomes nearly pure (ground for $\beta\to+\infty$ or excited for $\beta\to-\infty$), and $F_{\beta\beta}^{(0)}$ is suppressed.
The overall magnitude of $F_{\beta\beta}^{(0)}$ remains of order unity and is essentially independent of $N$, reflecting that temperature information is encoded in the single-atom thermal visibility, while the NOON super-resolution mainly boosts phase sensitivity.
For $N=10$ the thermal information is modulated by multiple narrow fringes along $x$, aligned with the phase-sensitive lobes of $F_{xx}^{(0)}$.

\begin{figure}[ht!]
    \centering
    \includegraphics[width=0.48\textwidth]{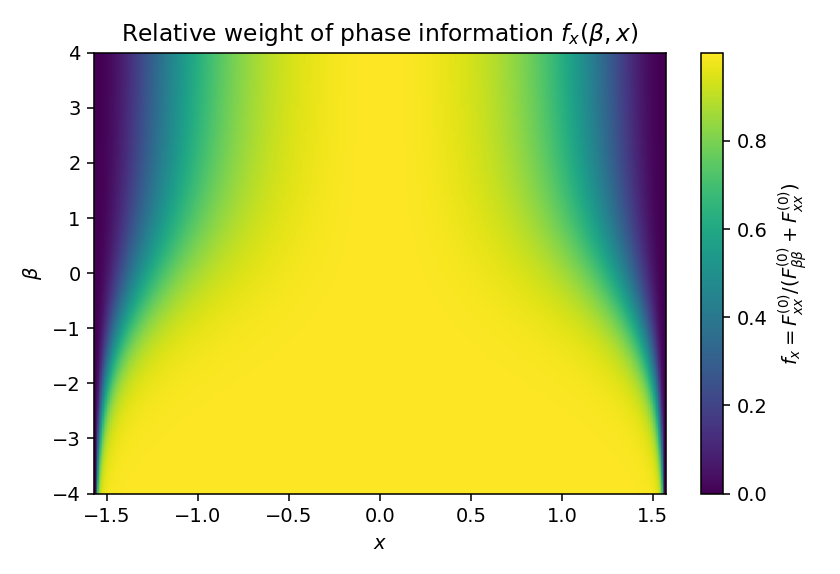}\hfill
    \includegraphics[width=0.48\textwidth]{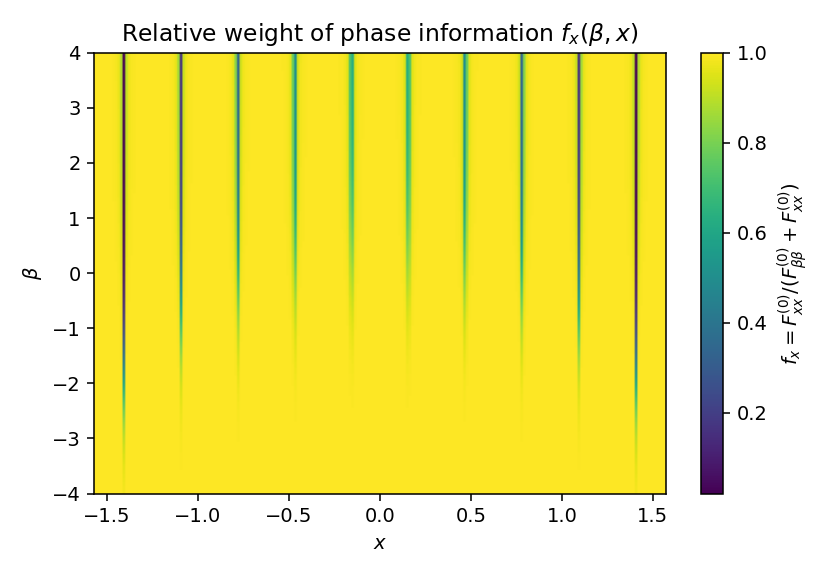}
    \caption{%
        Coordinate-dependent relative-response diagnostic
        $f_x(\beta,x) = F_{xx}^{(0)}/(F_{\beta\beta}^{(0)}+F_{xx}^{(0)})$ for NOON probes with
        (a)~$N=1$ and (b)~$N=10$ photons.
        Yellow regions ($f_x\simeq1$) correspond to regimes dominated by dispersive-phase estimation, most prominently in the inverted (negative-$\beta$) sector where the visibility is high.
        Darker regions highlight thermometric bands where $F_{\beta\beta}^{(0)}$ becomes comparable to $F_{xx}^{(0)}$.
    }
    \label{fig:fx_landscape}
\end{figure}

As a coordinate-dependent visualization of the relative response to phase and temperature, we define the fraction
\begin{equation}
    f_x(\beta,x) =
    \frac{F_{xx}^{(0)}(\beta,x)}{F_{\beta\beta}^{(0)}(\beta,x) + F_{xx}^{(0)}(\beta,x)},
\end{equation}
which ranges from $0$ (thermometric response dominates this single-setting diagnostic) to $1$ (phase response dominates this single-setting diagnostic).  Because $f_x$ depends on the chosen dimensionless parameterization, it is not an invariant multiparameter precision measure.
Figure~\ref{fig:fx_landscape} shows $f_x(\beta,x)$ for NOON probes with $N=1$ [panel~(a)] and $N=10$ [panel~(b)].
For $N=1$, the interferometer behaves predominantly as a phase sensor in the inverted regime (more negative $\beta$), while an intermediate band of $\beta$ exists where $F_{\beta\beta}^{(0)}$ becomes comparable to $F_{xx}^{(0)}$ and the device effectively acts as a thermometer.
For $N=10$ the phase sensitivity is further enhanced: $f_x$ is close to unity over a broad region of the $(\beta,x)$ plane, and the pronounced thermometric regions appear as thin horizontal stripes at phase values where $F_{xx}^{(0)}$ is suppressed by destructive interference.

\begin{figure}[ht!]
    \centering
    \includegraphics[width=0.48\textwidth]{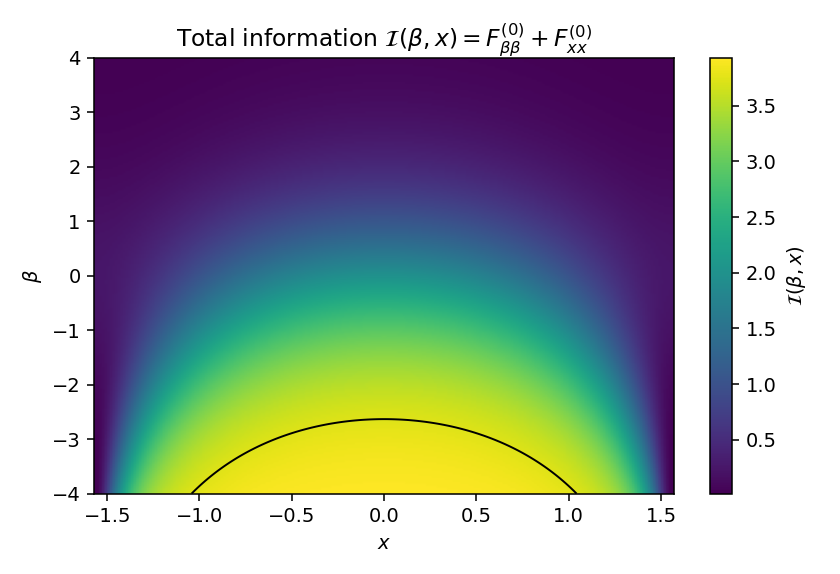}\hfill
    \includegraphics[width=0.48\textwidth]{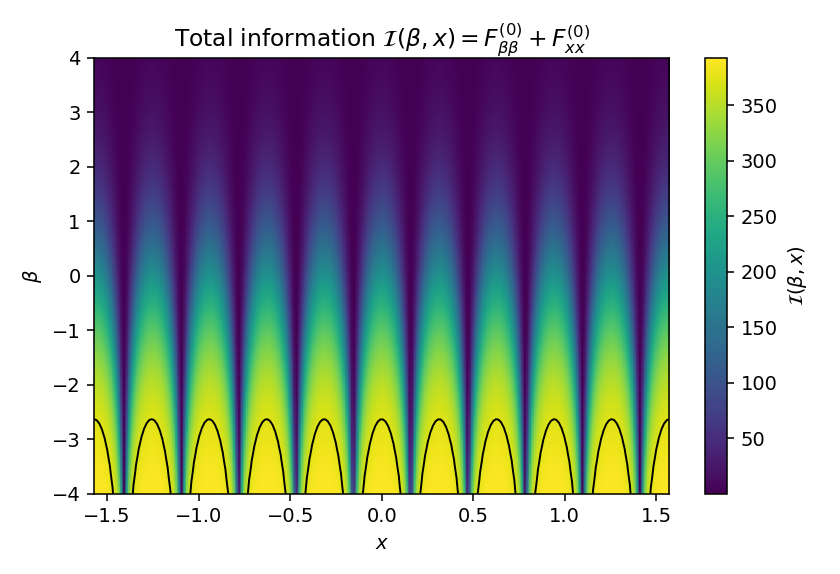}
    \caption{%
        Trace diagnostic $\mathcal{I}(\beta,x) = F_{\beta\beta}^{(0)}(\beta,x)+F_{xx}^{(0)}(\beta,x)$ for one binary photon-counting setting and NOON probes with
        (a)~$N=1$ and (b)~$N=10$ photons.
        In panel~(a) the black contour encloses the region where $\mathcal{I}\ge(1-\delta)\,\mathcal{I}_{\max}$ with $\delta=0.05$, corresponding to a high-response plateau of operating points that retain at least $95\%$ of the maximal information.
        For $N=10$ [panel~(b)] the information scale is enhanced roughly as $N^2$ and multiple equivalent near-maximal points for this single-setting trace diagnostic emerge along $x$, one in each interference lobe, as indicated by the black contour.
    }
    \label{fig:Itot_landscape}
\end{figure}

A second coordinate-dependent single-setting diagnostic is the trace of the photon-counting FIM,
\begin{equation}
    \mathcal{I}(\beta,x)
    = \mathrm{Tr}[F^{(0)}(\beta,x)] = F_{\beta\beta}^{(0)}(\beta,x) + F_{xx}^{(0)}(\beta,x).
\end{equation}
Figure~\ref{fig:Itot_landscape} displays $\mathcal{I}(\beta,x)$ as a function of $(\beta,x)$ for NOON probes with $N=1$ [panel~(a)] and $N=10$ [panel~(b)].
For $N=1$ the information is maximal in regions where the fringe slope is large and the visibility is appreciable, typically on the negative-$\beta$ side where the excited branch is significantly populated.
The black contour marks the region where
$\mathcal{I}(\beta,x)\ge(1-\delta)\,\mathcal{I}_{\max}$ with $\delta=0.05$, thus identifying a high-response plateau of operating points for this scalar diagnostic.  It should not be interpreted as an optimum for joint estimation unless a cost matrix and a multi-setting protocol have also been specified.
For $N=10$ the landscape is dominated by a sequence of narrow lobes along $x$, inherited from the super-resolved interference fringes of $F_{xx}^{(0)}$.
The overall scale of $\mathcal{I}$ grows approximately as $N^2$, and several equivalent near-maximal points for this single-setting trace diagnostic appear, one at the center of each high-visibility lobe, as highlighted by the black contour.

\subsection{Multi-setting identifiability and estimator benchmark}
\label{subsec:multisetting_results}

\begin{figure*}[t]
    \centering
    \includegraphics[width=\textwidth]{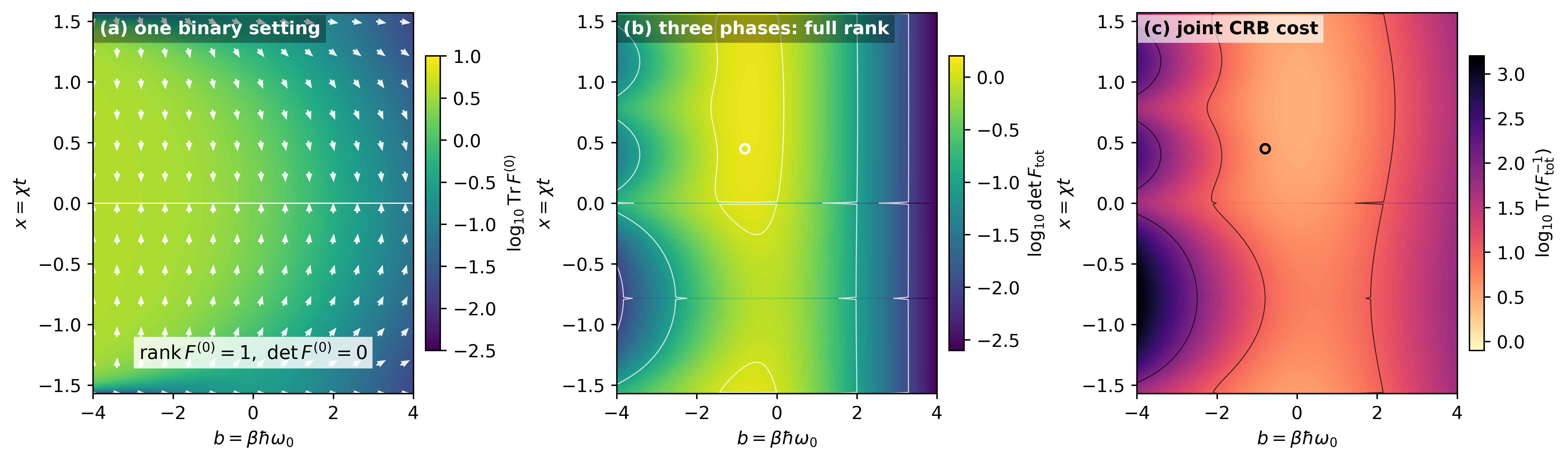}
    \caption{Local identifiability from controlled analysis phases.  (a) One binary photon-counting setting has nonzero sensitivity, shown by $\log_{10}\mathrm{Tr}\,F^{(0)}$, but only along one local score direction, shown by the arrows; hence $\operatorname{rank}F^{(0)}=1$ and $\det F^{(0)}=0$.  (b) Combining three known phases $\phi_j\in\{0,\pi/2,\pi\}$ yields a full-rank total Fisher matrix over broad regions of the parameter plane.  White contours mark representative determinant levels, and the open marker indicates the operating point used in Fig.~\ref{fig:mle_benchmark}.  (c) The corresponding weighted cost for $W=\mathbb{I}$, $\mathrm{Tr}(F_{\rm tot}^{-1})$, identifies favorable operating windows.  Here $b=\beta\hbar\omega_0$ and equal shot allocation is used, so the plotted cost rescales inversely with the total number of repetitions.}
    \label{fig:multisetting_identifiability}
\end{figure*}

The single-setting maps above are useful sensitivity diagnostics, but they do not by themselves constitute a locally invertible two-parameter protocol.  Figure~\ref{fig:multisetting_identifiability} makes this distinction explicit.  Panel~(a) shows that a one-setting binary measurement can carry nonzero information while still being one-dimensional: the color scale gives $\log_{10}\mathrm{Tr}\,F^{(0)}$, and the arrows show the local score direction in the $(b,x)$ plane.  Because the matrix has the outer-product form in Eq.~\eqref{eq:FIM_def_setting_j}, it has rank one and $\det F^{(0)}=0$ at every regular point.  Panels~(b) and (c) show the result of combining three controlled analysis phases, $\phi_j\in\{0,\pi/2,\pi\}$, with equal shot allocation.  The summed matrix $F_{\rm tot}$ is full rank over broad operating regions, and the weighted cost $\mathrm{Tr}(F_{\rm tot}^{-1})$ is finite wherever the score vectors are sufficiently non-collinear.  Singular or poorly conditioned points can still occur when all score vectors are nearly parallel or when both $\mathcal V$ and $|\mathcal V'|$ suppress the response; these points should be avoided when choosing an operating region.  This is the operational Fisher matrix that should be used for simultaneous estimation of $(\beta,x)$.

\begin{figure}[t]
    \centering
    \includegraphics[width=0.48\textwidth]{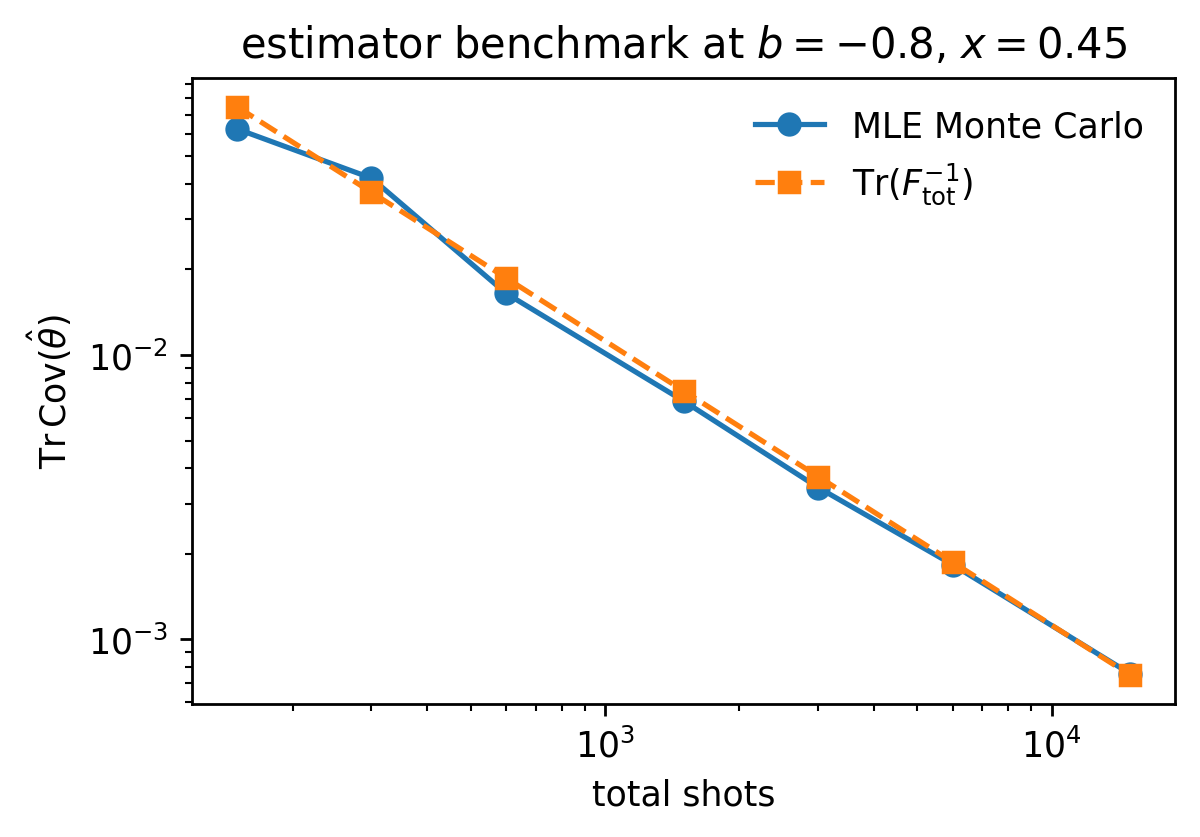}
    \caption{Synthetic estimator benchmark for the multi-setting protocol at $b=-0.8$ and $x=0.45$, using $\phi_j\in\{0,\pi/2,\pi\}$.  The empirical covariance of the maximum-likelihood estimator follows the predicted Cram\'{e}r--Rao scaling from $F_{\rm tot}^{-1}$.  Deviations at small sample size are expected from finite-sample bias and the nonlinearity of the inverse problem.}
    \label{fig:mle_benchmark}
\end{figure}

The MLE benchmark in Fig.~\ref{fig:mle_benchmark} was generated from the likelihood in Eq.~\eqref{eq:multisetting_likelihood} using the same three phases, equal shot allocation, bounded parameters in the plotted domain, and repeated synthetic binomial data sets.  For each data set the log-likelihood was maximized over $(\beta,x)$, and the empirical covariance of the resulting estimates was compared with $F_{\rm tot}^{-1}$ at the true operating point.  The benchmark is therefore an estimator-level validation of the synthetic multi-setting likelihood, not a validation of the IBM hardware run.

To verify that this multi-setting Fisher matrix corresponds to an explicit statistical procedure, we performed a synthetic estimator benchmark using the likelihood in Eq.~\eqref{eq:multisetting_likelihood}.  For a representative operating point $b=-0.8$, $x=0.45$, and the same three analysis phases, we generated binomial count records, applied the maximum-likelihood estimator of Eq.~\eqref{eq:mle_estimator}, and compared the empirical covariance with the predicted Cram\'{e}r--Rao cost.  Figure~\ref{fig:mle_benchmark} shows the expected $1/\mu$ scaling and close approach to $\mathrm{Tr}(F_{\rm tot}^{-1})$ as the number of shots increases.  This numerical experiment is not a substitute for hardware-level blind sensing, but it closes the statistical loop between the analytical Fisher matrix, the count record, and an operational estimator.

\subsection{Benchmarking the photon-counting FIM with circuit simulations}

\begin{figure}[ht!]
    \centering
    \includegraphics[width=0.45\textwidth]{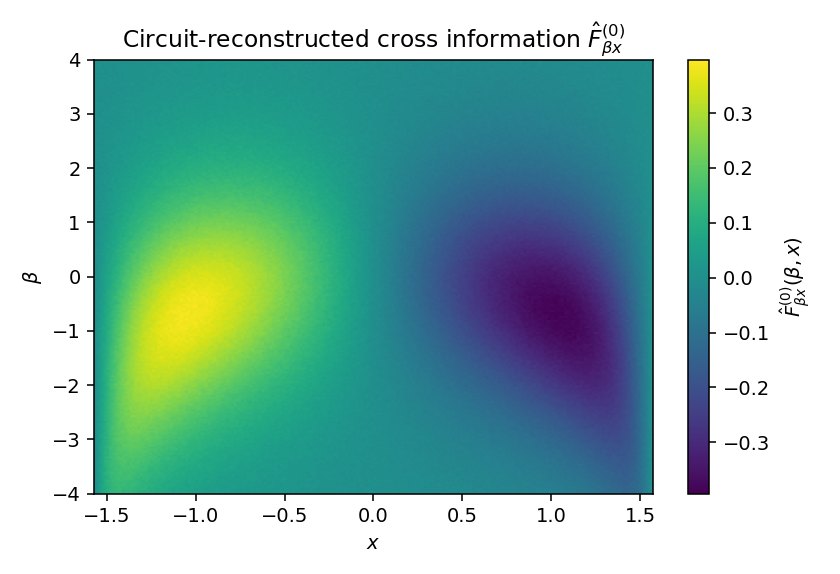}\hfill
    \includegraphics[width=0.45\textwidth]{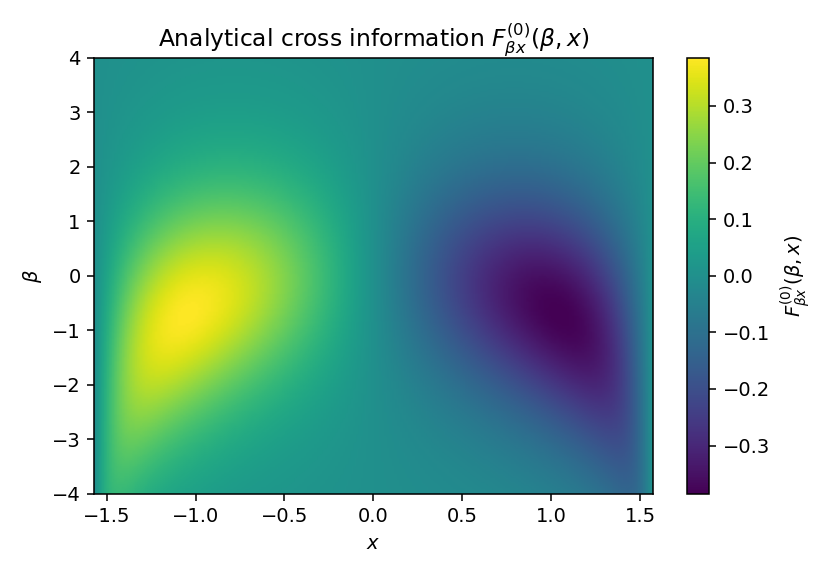}
    \caption{%
        Cross Fisher information $F_{\beta x}^{(0)}(\beta,x)$ from circuit simulation (top) and from the analytical expression (bottom).
        The two 3D surfaces coincide within statistical fluctuations, confirming that the digital implementation reproduces the continuous interferometric model and its joint dependence on $(\beta,x)$.
    }
    \label{fig:QFI_comparison_x}
\end{figure}

To benchmark the quantum-circuit implementation, we compare the Fisher information extracted from the simulated statistics with the analytical prediction for a representative off-diagonal element of the photon-counting FIM.
Figure~\ref{fig:QFI_comparison_x} shows the color map of the cross Fisher information $F_{\beta x}^{(0)}(\beta,x)$ obtained from the circuit (top) and from the analytical formula (bottom).
The agreement is excellent over the whole parameter region considered: both the magnitude and the shape of the surface are faithfully reproduced by the simulation.
Small discrepancies at the edges of the domain are consistent with finite-sampling fluctuations due to the large imbalance between the two outcomes when $P_0$ is very close to $0$ or $1$.

For completeness, we have also computed analogous surfaces for the diagonal elements $F_{xx}^{(0)}$ and $F_{\beta\beta}^{(0)}$.
They display the same level of agreement between simulation and theory as in Fig.~\ref{fig:QFI_comparison_x} and are therefore not shown here.

% -----------------------------------------------------------------
\subsection{Hardware emulation on IBM Quantum hardware}
\label{sec:experiment_ibm}

We executed the circuit of Fig.~\ref{fig:MZ_circuit} on IBM Quantum hardware to emulate the likelihood and its ideal-score Fisher diagnostics under device noise and finite sampling. The transpiled circuits were run on the backend \texttt{ibm\_torino} using $\mu=10\,000$ shots per grid point. Qiskit optimization level~1 was used, and no error-mitigation procedure was applied.

\paragraph{Hardware execution and data acquisition.}
For each point of a uniformly spaced $20\times 20$ grid in $(b,x)$ with $b\in[-4,4]$ and $x\in[-\pi/2,\pi/2]$, we compiled the four-qubit circuit of Fig.~\ref{fig:MZ_circuit} to the native gate set of the chosen IBM backend. The measurement consists of a projective readout of the light qubit (encoding the interferometer output port) in the computational basis. From the raw counts $n_0$ and $n_1$, we estimate the empirical frequency $\hat P_0(\beta,x)=n_0/\mu$, which converges to the ideal probability $P_0(\beta,x)$ in the limit of large $\mu$. In practice, device noise and finite sampling reduce the fringe contrast and introduce small asymmetries across the parameter domain.

\paragraph{Reconstruction of the Fisher-information matrix.}
Given the two-outcome statistics, we construct ideal-score Fisher diagnostics using the same analytical score functions as in the ideal likelihood, but with the probabilities in the denominator replaced by the experimental frequencies $\hat P_0$.  These quantities reveal how hardware contrast loss distorts the ideal information landscape.  They are not, by themselves, the Fisher matrix of the unknown noisy hardware likelihood, because that would require the derivatives $\partial_a P_0^{\rm hw}$ of a calibrated noise model.

\begin{figure*}[t]
  \centering
  \subfloat[]{\includegraphics[width=0.48\textwidth]{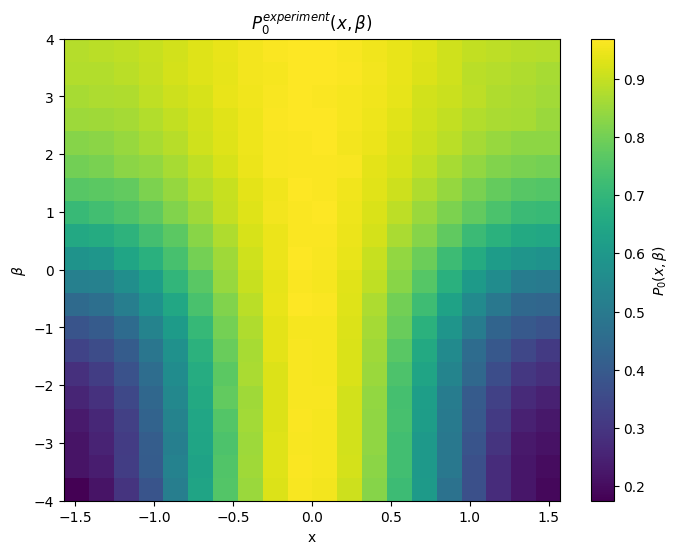}}\hfill
  \subfloat[]{\includegraphics[width=0.48\textwidth]{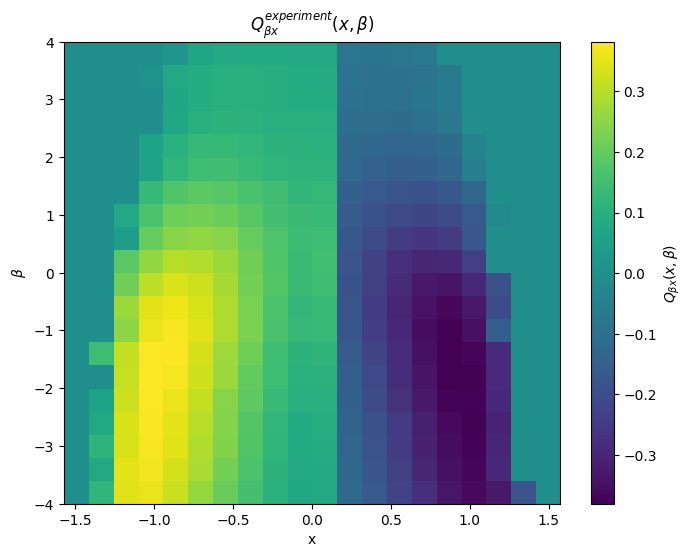}}\\[-0.3em]
  \subfloat[]{\includegraphics[width=0.48\textwidth]{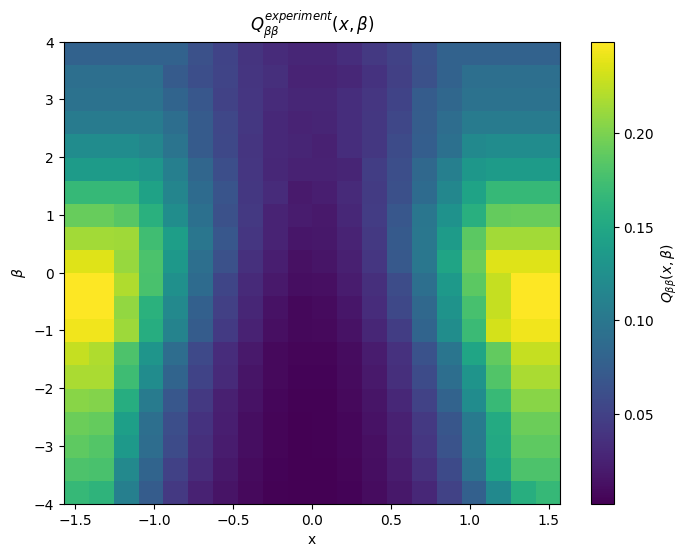}}\hfill
  \subfloat[]{\includegraphics[width=0.48\textwidth]{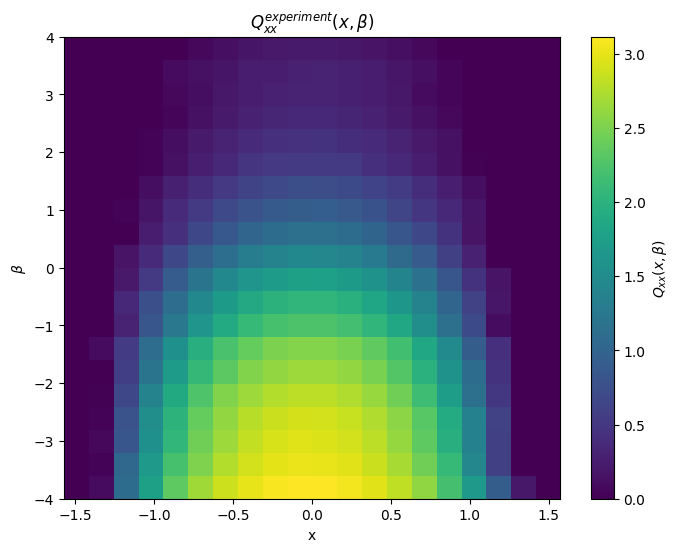}}
  \caption{Hardware-emulation reconstruction on IBM Quantum hardware. (a) Measured interference probability $\hat P_{0}(b,x)$. (b) Hardware-distorted ideal-score cross diagnostic $\hat F_{\beta x}(b,x)$. (c) Hardware-distorted ideal-score diagnostic $\hat F_{\beta\beta}(b,x)$. (d) Hardware-distorted ideal-score diagnostic $\hat F_{xx}(b,x)$. These panels evaluate ideal score functions on hardware frequencies; they show qualitative contrast loss but should not be read as a fully calibrated Fisher matrix of the noisy hardware likelihood.}
  \label{fig:exp_landscapes}
\end{figure*}

\paragraph{Experimental landscapes.}
Figure~\ref{fig:exp_landscapes} summarizes the experimentally reconstructed surfaces. Panel~(a) shows the measured interference probability $\hat P_0(b,x)$, which retains the expected periodic dependence on $x$ but displays device-induced contrast distortion relative to the ideal prediction. Here ``visibility'' denotes the operational contrast $\mathcal{V}_{\rm meas}\equiv(P_{\max}-P_{\min})/(P_{\max}+P_{\min})$ extracted from $\hat P_0(b,x)$. In the unitary limit, $\mathcal{V}_{\rm meas}=\mathcal{V}(\beta)$ [Eq.~\eqref{eq:visibility}]. In the phenomenological model used below, device noise and SPAM errors pull the inferred visibility toward the baseline $\mathcal{V}(0)=1/3$, thereby reducing its distance from the infinite-temperature reference rather than necessarily decreasing its numerical value.
Panels (b)--(d) show these hardware-distorted ideal-score Fisher entries. Despite the reduced absolute magnitudes, the hardware run reproduces the main qualitative features discussed in Sec.~\ref{sec:results}: (i) $\hat F_{\beta\beta}$ is concentrated in the intermediate region where the visibility changes most rapidly with $\beta$; (ii) $\hat F_{xx}$ is largest in regions where the fringe slope is maximal; and (iii) the cross term $\hat F_{\beta x}$ changes sign across $x=0$, reflecting the odd symmetry associated with $\partial_x P_0\propto\sin(2Nx)$.

\paragraph{Error analysis and systematic bias.}
Bootstrap error bars for the ideal-score Fisher diagnostics $\hat{F}_{ij}$ were estimated by resampling the hardware shot counts. Away from singular points, near-deterministic probabilities, and zero crossings of the cross term, the uncertainties are of the expected order $1/\sqrt{\mu}$. The hardware-emulation surfaces in Fig.~\ref{fig:exp_landscapes} represent raw data without error mitigation or noise fitting.
To interpret the deviations, we inferred point estimates $(\hat\beta,\hat x)$ from the fringe data (Table~\ref{tab:exp_estimates}). We observe a systematic bias consistent with a \emph{contrast shrinkage} mechanism: the measured fringe visibility is pulled toward a baseline value close to $\mathcal{V}(0)=1/3$ (infinite-temperature reference), i.e.,
\begin{equation}
  \mathcal{V}_{\mathrm{meas}}
  \simeq \kappa\,\mathcal{V}(\beta_{\mathrm{true}}) + (1-\kappa)\,\mathcal{V}(0),
  \qquad 0<\kappa<1,
\end{equation}
which is a minimal effective model for SPAM errors and unmitigated decoherence that partially randomize the output statistics. 
The parameter $\kappa$ is phenomenological rather than the result of a microscopic noise model. This interpolation captures the observed shrinkage of the inferred visibility toward $\mathcal{V}(0)$ while preserving the analytical dependence of the likelihood on the thermal visibility.
Since the estimator $\hat\beta=\mathcal{V}^{-1}(\mathcal{V}_{\mathrm{meas}})$ is constructed by inverting the monotonic visibility map (Appendix~\ref{app:bias_model}), this shrinkage biases the estimate toward $\beta\simeq 0$, yielding $|\hat\beta|<|\beta_{\mathrm{true}}|$ for both signs of $\beta$.
This behavior is consistent with contrast-shrinking device noise, although the present phenomenological model does not identify or separate the underlying microscopic channels (see Appendix~\ref{app:bias_model}).

% ---------------------- FIGURE (2x2) ----------------------------

% ---------------------- TABLE (REPRESENTATIVE POINTS) -----------
\begin{table}[t]
\caption{Representative hardware-emulation point estimates (IBM hardware) and corresponding ideal-simulator values. Values were extracted from the $20\times 20$ grid used to build Fig.~\ref{fig:exp_landscapes} by fitting the measured fringe versus $x$ at fixed nominal $\beta$ to the visibility form in Eq.~\eqref{eq:probabilities_Vstd} and then applying the inverse map in Eq.~\eqref{eq:beta_from_V}; the listed $\hat x$ is the fitted phase coordinate. Statistical intervals are not tabulated here and should be reported with the raw hardware-count release.}
\label{tab:exp_estimates}
\begin{ruledtabular}
\begin{tabular}{cc|ccc|ccc}
$\beta$ & $x$ & $\hat\beta_{\rm exp}$ & $\hat x_{\rm exp}$ & $\hat P_{0,\rm exp}$ & $\hat\beta_{\rm sim}$ & $\hat x_{\rm sim}$ & $\hat P_{0,\rm sim}$ \\
\hline
-4.0 & $-\pi/2$ & -1.308 & -1.571 & 0.175 & -3.842 & -1.571 & 0.017 \\
-4.0 & -0.083   & -1.308 & -0.222 & 0.962 & -3.842 & -0.085 & 0.993 \\
-4.0 & $+\pi/2$ & -1.308 & 1.571 & 0.193  & -3.842 &  1.570 & 0.021 \\
\hline
 0.211 & $-\pi/2$ & 0.467 & -1.571 & 0.584 &  0.084 & -1.331 & 0.548 \\
 0.211 & -0.083   & 0.467 & -0.299 & 0.966 &  0.084 & -0.046 & 0.999 \\
 0.211 & $+\pi/2$ & 0.467 & 1.571  & 0.581 &  0.084 &  1.571 & 0.520 \\
\hline
 4.0 & $-\pi/2$ & 2.345 & -1.571 & 0.88  &  3.749 & -1.565 & 0.977 \\
 4.0 & -0.083   & 2.345 & -0.657 & 0.967 &  3.749 &  0.000 & 1.000 \\
 4.0 & $+\pi/2$ & 2.345 & 1.571  & 0.88  &  3.749 &  1.271 & 0.979 \\
\end{tabular}
\end{ruledtabular}
\end{table}

\subsection{Interpretation and physical implications}

The combined analytical and numerical results confirm that the interferometer acts as a dual probe: it can function as a high-precision phase sensor in the \emph{inverted} (negative-$\beta$) regime or as a thermometer in an intermediate band of $\beta$ where the visibility varies most rapidly.
For large negative $\beta$, the atomic ancilla is almost pure in the excited state, the visibility approaches unity, and the phase information $F_{xx}^{(0)}$ dominates with Heisenberg-like scaling in $N$.
For large positive $\beta$, the ancilla is almost pure in the ground state, $\mathcal{V}(\beta)\to 0$, and both $F_{xx}^{(0)}$ and $F_{\beta\beta}^{(0)}$ are suppressed because no dispersive phase is imprinted and the output statistics become nearly deterministic ($P_0\to1$).
For intermediate $\beta$, thermal mixing makes $|\mathcal{V}'(\beta)|$ appreciable, leading to enhanced temperature sensitivity encoded in $F_{\beta\beta}^{(0)}$.
Thus, the same interferometric architecture can operate in distinct metrological modes, governed by the single control parameter $\beta$ through $\mathcal{V}(\beta)$.

\subsection{Tuning and inverted populations}

\paragraph{Resonance Tuning ($\omega_0$).}
The trade-off between thermometer and phase-sensor regimes is governed by the energy scale $\hbar\omega_0$. The temperature sensitivity peaks when the visibility slope is maximal, i.e., when $\beta$ is such that $\mathcal{V}(\beta)\big[1-\mathcal{V}(\beta)\big]$ is appreciable. Consequently, the sensor is tunable: by adjusting the atomic splitting $\omega_0$ (e.g., via Zeeman or Stark shifts), one can shift the effective \emph{thermometer window} to the target temperature range. We note that inhomogeneous broadening (a distribution of $\omega_0$ across an ensemble) would act as an additional dephasing channel, averaging out $\mathcal{V}(\beta)$ and reducing the peak Fisher information.

\paragraph{Negative Temperatures.}
Our analysis extends naturally to negative temperatures ($\beta < 0$), relevant for population-inverted systems.
Experimentally, such states can be prepared in the thermal ancilla by engineered inversion protocols (e.g., a high-fidelity $\pi$-pulse applied to an initially equilibrated two-level system, followed by controlled mixing).
Within our operational definition, inversion corresponds to $\mathcal{V}(\beta)>\mathcal{V}(0)=1/3$, and it manifests as an enhancement of fringe contrast (deeper minima) rather than as a $\pi$ phase flip.
The cross-correlation term $F_{\beta x}$ changes sign across $x=0$ because it is proportional to $\sin(2Nx)$, while its $\beta$-dependence is governed by the monotonic factor $\mathcal{V}(\beta)\big[1-\mathcal{V}(\beta)\big]$.
Hence, the metrological signature of inversion in this model is encoded in the \emph{contrast level} (and thus in $\mathcal{V}$), not in an antisymmetry under $\beta\to-\beta$.

%%%%%%%%%%%%%%%%%%%%%%%%%%%%%%%%%%%%%%%%%%%%
\section{Robustness analysis and experimental limitations}
\label{sec:robustness_analysis}
The transition from idealized photon-counting models to practical sensing is controlled by decoherence, loss, and state-preparation-and-measurement errors.  The clean visibility picture of Sec.~\ref{sec:model} is therefore best regarded as a baseline likelihood model.  In this section we assess how amplitude damping (AD) and phase damping (PD) degrade the information carried by several standard probe families.  We use Fisher-information susceptibility $\chi_F$~\cite{albarelli2024fisher} as a local diagnostic of fragility.  The goal is not to prove a universal ordering of all possible probes, but to compare NOON, cat, and squeezed probes under common noise assumptions and to interpret the contrast shrinkage observed in the IBM hardware-emulation data of Sec.~\ref{sec:experiment_ibm}.

\subsection{Decoherence models and noise channels}
\label{subsec:decoherence_models}

In any physical realization of the MZI, the probe experiences simultaneous energy relaxation (amplitude damping) and dephasing (phase damping). We model this non-unitary evolution via the Lindblad master equation~\cite{campaioli2024quantum}:

\begin{equation}
\frac{d\rho}{dt} = -i[\hat{H}(\boldsymbol{\theta}), \rho] + \gamma_{\text{AD}} \mathcal{D}[\hat{a}]\rho + \gamma_{\text{PD}} \mathcal{D}[\hat{a}^\dagger \hat{a}]\rho,
\end{equation}
where $\mathcal{D}[\hat{L}]\rho = \hat{L}\rho\hat{L}^\dagger - \frac{1}{2}\{\hat{L}^\dagger\hat{L}, \rho\}$ is the standard dissipator. To map these continuous-time dynamics onto the discrete parameter space of our results (Fig.~\ref{fig:decoherence_landscape}), we define the dimensionless \emph{transmission efficiency} $\eta$ and \emph{dephasing strength} $\gamma$ as:
\begin{equation}
    \eta \equiv e^{-\gamma_{\text{AD}} t}, \quad \gamma \equiv \frac{\gamma_{\text{PD}} t}{2},
\end{equation}
where $t$ is the total interaction time. Physically, $\eta$ represents the probability that a single photon survives the passage through the interferometer arms (where $1-\eta$ is the loss probability), while $\gamma$ sets the accumulated phase-diffusion variance in the NOON coherence.

Under these channels, the ideal thermal visibility $\mathcal{V}(\beta)$ derived in Eq.~\eqref{eq:visibility} degrades into an effective visibility $\mathcal{V}_{\text{eff}}$.  The expression below assumes equal loss in the two arms and an unconditional single-port visibility measurement; a postselected coincidence protocol would instead acquire an additional survival factor.  For an $N$-photon NOON state, which requires the coherence of the maximally separated Fock superposition, this degradation scales as (see Appendix~\ref{app:master_equation} for the explicit derivation from the Lindblad dynamics):
\begin{equation}
    \mathcal{V}_{\text{eff}}(\beta,\eta,\gamma) = \eta^{N/2} e^{-\gamma N^2} \mathcal{V}(\beta).
\end{equation}

Note that the scaling with $N$ highlights the extreme susceptibility of this state: the term $e^{-\gamma N^2}$ reflects the quadratic enhancement of phase noise sensitivity, while $\eta^{N/2}$ (or $\eta^N$ for coincidence detection) captures the exponential probability of avoiding photon loss. This definition of $\eta$ provides the rigorous physical basis for the x-axis in Fig.~\ref{fig:decoherence_landscape}, allowing us to directly correlate the dynamical decay rates with the metrological performance limits.

\subsection{Probe-family robustness under joint noise}
\label{subsec:probe_family_robustness}

\begin{figure*}[t]
    \centering
    \includegraphics[width=\textwidth]{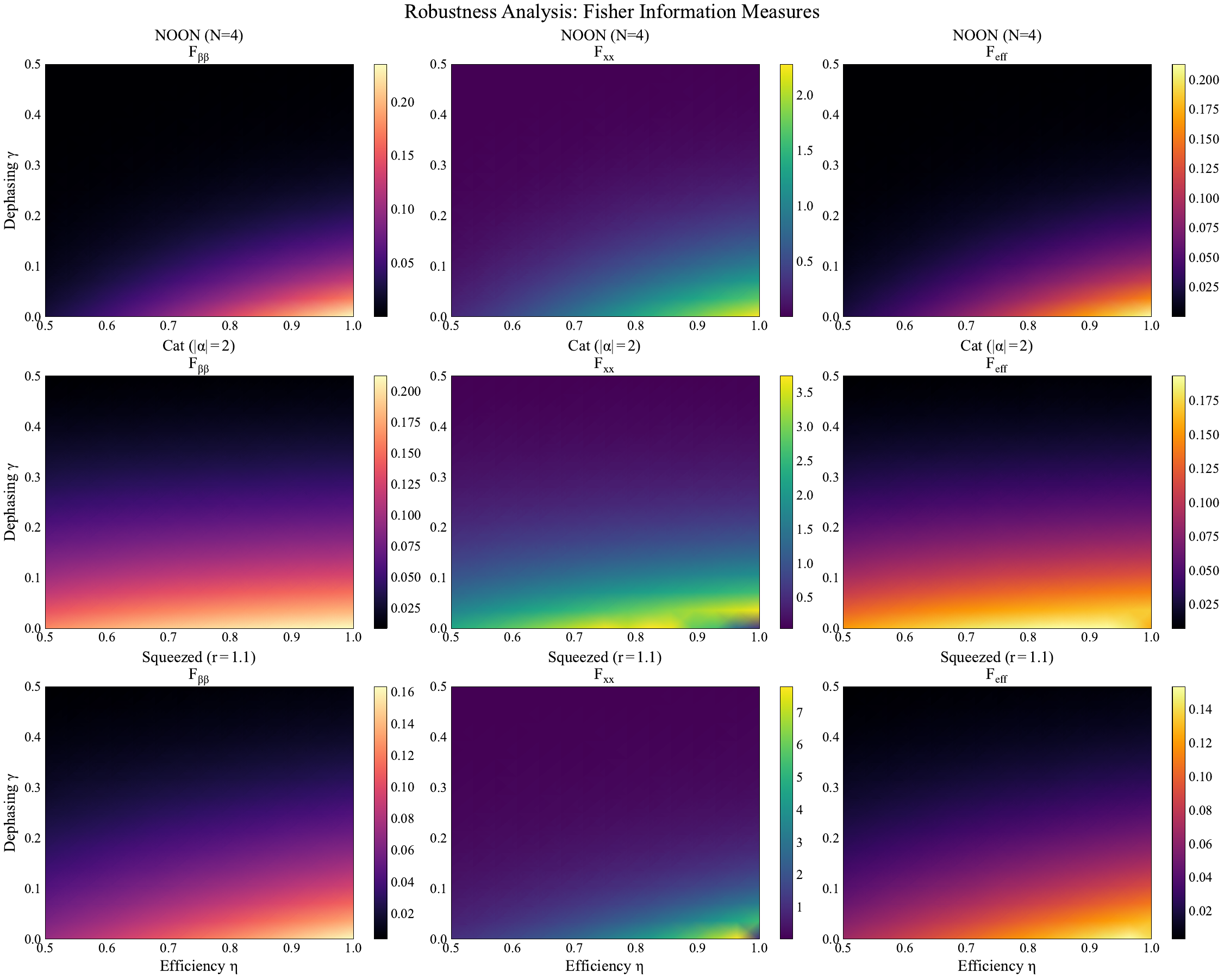}
    \caption{\textbf{Decoherence landscape and probe-family robustness.}
    Evaluation of the quantum Fisher information (QFI) elements $F_{\beta\beta}$ (temperature, left column), $F_{xx}$ (phase, middle column), and the scalar robustness diagnostic $F_{\text{eff}}$ defined in Eq.~\eqref{eq:Feff_def} (right column) for representative NOON ($N=4$), cat ($|\alpha|=2$), and squeezed ($r=1.1$) states.
    The performance is mapped across the parameter space of transmission efficiency $\eta$ (x-axis) and dephasing rate $\gamma$ (y-axis), evaluated at the representative operating point $\beta=0.5$ and $x=\pi/4$.
    (Top row) \textbf{NOON states} show high peak sensitivity but lose useful information rapidly away from the high-transmission, low-dephasing corner in this representative parameter set.
    (Middle row) \textbf{Cat states} offer a non-Gaussian compromise in the plotted regime.
    (Bottom row) \textbf{Squeezed states} maintain high $F_{\text{eff}}$ over a larger area of this specific noise landscape, showing why they are candidates for open-system sensing under these assumptions.}
    \label{fig:decoherence_landscape}
\end{figure*}

\begin{figure*}[t]
    \centering
    \includegraphics[width=\textwidth]{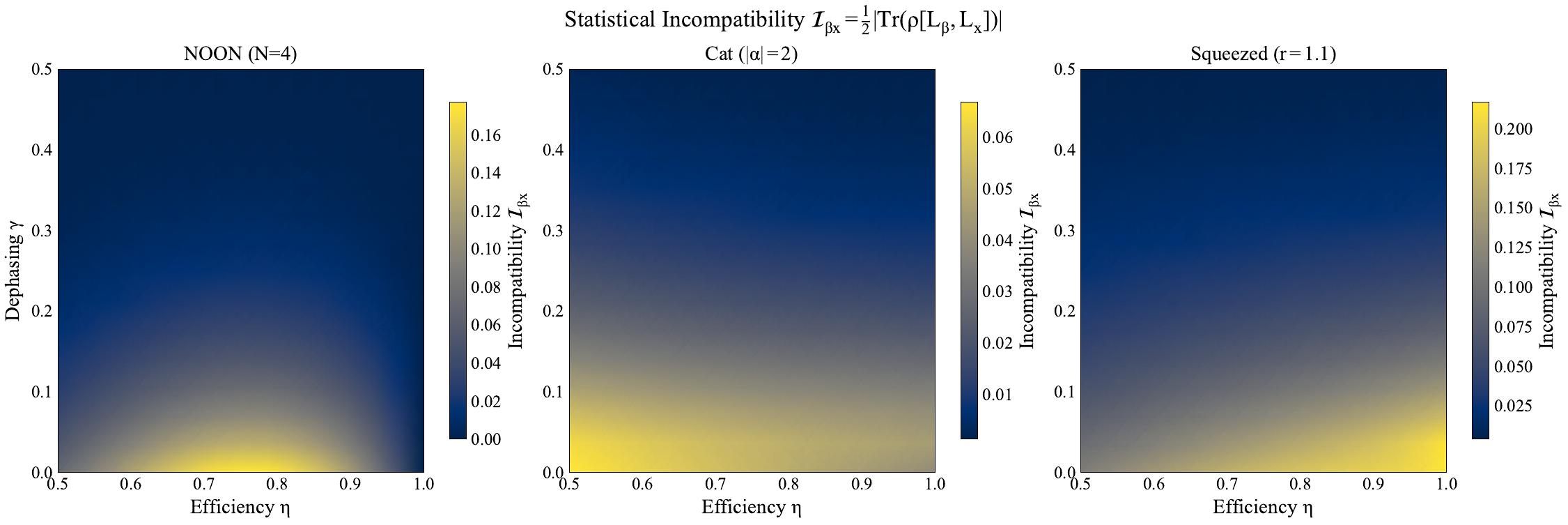}
    \caption{\textbf{Statistical Incompatibility $\mathcal{I}_{\beta x}$.}
    The magnitude of the commutator expectation value $\mathcal{I}_{\beta x} = \frac{1}{2}|\mathrm{Tr}(\rho [L_\beta, L_x])|$ for (a) NOON, (b) Cat, and (c) Squeezed states under joint amplitude and phase damping.
    Brighter regions indicate higher incompatibility and therefore a stronger obstruction to simultaneous saturation of the single-parameter symmetric-logarithmic-derivative bounds for temperature and phase.
    Notably, \textbf{Squeezed states} (c) retain a \emph{ridge} of high incompatibility in the low-noise regime, indicating that their robustness can come with a more demanding simultaneous-readout requirement.}
    \label{fig:incompatibility}
\end{figure*}

To quantify the impact of these channels, we compare three standard probe families under the same noise model and operating point.  The comparison uses representative few-photon-scale probes: a NOON state with $N=4$, a cat state with $|\alpha|=2$ (mean photon number of order four), and a two-mode squeezed state with $r=1.1$.  This is an approximate operating-regime match, not a strict equality of preparation cost, ideal QFI, or total optical energy.  The scalar plotted in the right column is
\begin{equation}
    F_{\rm eff}=
    \begin{cases}
    \left[\mathrm{Tr}\,F^{-1}\right]^{-1}, & \det F>0,\\[2mm]
    0, & \det F=0,
    \end{cases}
    \label{eq:Feff_def}
\end{equation}
with $F$ the two-parameter QFIM for $(\beta,x)$ in the corresponding noisy probe model and with dimensionless unit weights.  The Fisher-information surfaces in Fig.~\ref{fig:decoherence_landscape} and the incompatibility maps in Fig.~\ref{fig:incompatibility} should therefore be read as a controlled robustness comparison, not as a universal hierarchy over all possible measurements and encodings.

\textbf{NOON states: a loss-sensitive benchmark.}
NOON states achieve the ideal Heisenberg scaling $F_{xx}\propto N^2$, but the useful advantage is rapidly eroded by loss and dephasing.  In the representative $N=4$ landscape of Fig.~\ref{fig:decoherence_landscape}, the effective Fisher information $F_{\text{eff}}$ is concentrated close to the high-transmission, low-dephasing corner.  This behavior is consistent with the Fisher-information susceptibility derived in Appendix~\ref{app:susceptibility}, where the weak-loss fragility scales as $\chi_F\propto N^3$ for amplitude damping.  Thus, high-$N$ NOON states require exceptionally clean transmission and phase stability; outside that regime, lower-resource or more robust probes can provide a better operational choice~\cite{Afek2010,nielsen2023deterministic}.  Appendix~\ref{app:escher_connection} relates this susceptibility picture to the asymptotic bounds of Escher et al.~\cite{escher2011general}.

\textbf{Two-mode squeezed vacuum: Gaussian robustness.}
Two-mode squeezed vacuum (TMSV) states remain Gaussian under amplitude damping.  In the bottom row of Fig.~\ref{fig:decoherence_landscape}, the region of high $F_{\text{eff}}$ extends farther into the lossy regime than in the NOON case for the parameters shown.  Because photon loss transforms the covariance matrix continuously and degrades the effective squeezing rather than projecting onto a small loss-free subspace~\cite{Braunstein2005}, TMSV probes are attractive for steady-state sensing in lossy channels.  The trade-off is readout compatibility: as shown in Fig.~\ref{fig:incompatibility}(c), squeezed states exhibit a ridge of nonzero statistical incompatibility $\mathcal{I}_{\beta x}$ in the low-noise regime.  Extracting the full multiparameter advantage may therefore require measurement designs beyond the simple photon-counting readout used for the NOON protocol~\cite{ullah2025optimal, xia2023toward}. 

\textbf{Cat states: a non-Gaussian compromise.}
Schr\"{o}dinger cat states ($|\psi_{\text{cat}}\rangle \propto |\alpha\rangle + |-\alpha\rangle$) occupy an intermediate regime.  For the parameters shown, the middle row of Fig.~\ref{fig:decoherence_landscape} retains a larger preserved-information region under amplitude damping than the NOON benchmark.  The underlying reason is that a photon-loss event can map the even cat manifold into an odd cat manifold, so part of the phase-space structure survives when the parity information is tracked.  Figure~\ref{fig:incompatibility}(b) also indicates a smaller incompatibility diagnostic than the squeezed-state case in parts of the plotted operating region.  Cat states therefore provide a useful non-Gaussian compromise for transient, lossy regimes, especially when parity-sensitive readout is available~\cite{park2025quantum, mendonca2025information}.  Appendix~\ref{app:hierarchy_derivation} gives the susceptibility estimates used for the comparison with NOON and squeezed probes.

% ---------------- FIGURE 3: DECOHERENCE LANDSCAPE ----------------

The incompatibility diagnostic was evaluated from the noisy probe density matrix by diagonalizing $\rho=\sum_m\lambda_m|m\rangle\langle m|$ and constructing the SLDs as
\begin{equation}
    L_a=2\sum_{m,n:\lambda_m+\lambda_n>0}
    \frac{\langle m|\partial_a\rho|n\rangle}{\lambda_m+\lambda_n}
    |m\rangle\langle n|,\qquad a\in\{\beta,x\}.
    \label{eq:SLD_numerical}
\end{equation}
The plotted quantity is then $\mathcal I_{\beta x}=\frac12|\mathrm{Tr}[\rho(L_\beta L_x-L_xL_\beta)]|$.  Points with numerically negligible support in $\lambda_m+\lambda_n$ are omitted from the sum.

% ---------------- FIGURE 4: INCOMPATIBILITY ----------------

\subsection{Joint Discussion: Robustness vs. Incompatibility}
\label{subsec:joint_discussion}

The interplay between the Fisher-information landscapes (Fig.~\ref{fig:decoherence_landscape}) and the statistical incompatibility (Fig.~\ref{fig:incompatibility}) reveals a practical resource trade-off.

Figure~\ref{fig:decoherence_landscape} establishes a controlled comparison of structural robustness.  The NOON state is the most loss-sensitive of the three families shown: its high ideal response is concentrated near $\eta\simeq1$ and small dephasing.  The squeezed-state landscape shows a broader plateau in $F_{\text{eff}}$ for the same noise parameters, consistent with the continuous covariance-matrix degradation of Gaussian probes under loss.

Figure~\ref{fig:incompatibility} adds the corresponding readout constraint.  Robustness does not imply that the information is easy to access.  The squeezed-state case exhibits a nonzero incompatibility diagnostic $\mathcal{I}_{\beta x}$ [Fig.~\ref{fig:incompatibility}(c)], particularly in high-efficiency regimes, indicating that measurements optimized for temperature and phase need not coincide.  Experimentally approaching the theoretical limit therefore requires readout designs that are matched to the multiparameter objective, for example optimized Gaussian or generalized-dyne measurements when they are available.

Conversely, the NOON-state incompatibility diagnostic vanishes rapidly as noise increases [Fig.~\ref{fig:incompatibility}(a)], mainly because the state loses the coherence that carried the information.  The cat-state case remains an intermediate candidate: it can retain non-Gaussian phase-space structure in transient regimes while showing a smaller incompatibility diagnostic than the squeezed-state case in parts of the plotted domain.

\subsection{Platform-specific implementation constraints}
\label{subsec:platform_constraints}

The theoretical models derived above suggest the following platform-level constraints.

\subsubsection{Circuit QED (superconducting qubits)}
Two microwave resonator modes, or two effective paths encoded in a multimode circuit, serve as the MZI arms, while a transmon qubit serves as the thermal ancilla $\hat{\rho}_{\mathrm{A}}(\beta)$.

\begin{itemize}
    \item \textbf{Justification for the $N=1$ limit:}
    Our hardware emulation on \texttt{ibm\_torino} was deliberately restricted to the single-qubit probe limit ($N=1$). This choice is informed by the scaling analysis in Ref.~\cite{akamatsu2025fundamental}, where the coherence of a NOON state under phase damping contains the factor $\exp(-\gamma N^2 t)$.  By operating at $N=1$, the hardware run avoids the large-$N$ fragility regime and focuses on the contrast-shrinkage bias produced by relaxation, dephasing, and state-preparation-and-measurement errors.  As observed in Table~\ref{tab:exp_estimates} and detailed in Appendix~\ref{app:bias_model}, this shrinkage pulls the visibility-based estimate toward the infinite-temperature point $\beta=0$.

    \item \textbf{Favorable regime:}
    Despite $T_2$ limitations, this platform is well suited to transient thermometry protocols when the dispersive coupling is strong enough to keep the optimal interrogation time $t_{\text{opt}}$ short compared with the relevant dephasing and relaxation times.  Readout calibration remains essential.
\end{itemize}

\subsubsection{Integrated quantum photonics}
Waveguides implement the MZI paths, so propagation loss is a central constraint; representative reported losses span broad ranges depending on material platform and wavelength ($\alpha \sim 0.1-3$ dB/cm in the setting of Ref.~\cite{Bruck2020}).
\begin{itemize}
    \item \textbf{Constraint}: Unlike circuit-QED emulation, the dominant limitation is often transmission $\eta$.
    \item \textbf{Implication}: In the parameter set of Fig.~\ref{fig:decoherence_landscape}, squeezed probes retain useful information deeper into the lossy regime, whereas the NOON-based visibility $\mathcal{V}_{\text{eff}}$ vanishes rapidly \cite{qin2023unconditional}.
\end{itemize}

\subsubsection{Trapped-ion platforms}
The MZI modes $a$ and $b$ are mapped to collective motional modes (phonons) via Raman laser pulses \cite{Monroe2021}.
\begin{itemize}
    \item \textbf{Motional heating}: The environment adds phonons at a rate $\dot{\bar{n}}$ \cite{Brown2011}, producing a heating or thermalization channel that degrades the NOON interference $P_0(\beta,x)$.
    \item \textbf{Dicke-state alternative}: Symmetric Dicke states $|N/2, N/2\rangle_S$ are a relevant alternative for dispersive sensing because they can be less sensitive to local phase noise ($\gamma$) in Paul-trap settings \cite{he2025symmetric}.
\end{itemize}

\subsection{Error mitigation and resource optimization}
\label{subsec:error_mitigation}

To extend the protocol's operational range in the presence of noise, several mitigation strategies can be incorporated into the control layer:

\begin{itemize}
    \item \textbf{Adaptive preparation:} Real-time feedback allows the experimentalist to match state parameters---such as the squeezing degree $r$---to the transient noise levels \cite{quek2021adaptive}.
    \item \textbf{Dynamical decoupling:} Echo-type sequences in superconducting circuits can suppress low-frequency dephasing and extend the effective coherence time $T_2$ in favorable regimes \cite{Viola1999, Bylander2011}.
    \item \textbf{Bayesian inference:} Treating noise parameters $(\eta, \gamma)$ as nuisance variables within a joint estimation model allows for self-calibration. This can help correct the contrast-shrinkage bias observed in the IBM hardware emulation \cite{Granade2012}.
    \item \textbf{Variational compilation:} Variational algorithms can search for entangled probe preparations adapted to the measured noise structure of a device \cite{cimini2024variational}.
\end{itemize}

\subsection{Near-term extensions}
\label{subsec:future_directions}

The most important next step is a complete estimator-level experiment.  A future implementation should choose several non-collinear settings $\{\phi_j\}$, hide the true values of $(\beta,x)$ from the estimator, infer them from raw measurement records using maximum-likelihood or Bayesian methods, and compare the empirical covariance with $F_{\rm tot}^{-1}$ as the number of shots and photons is varied.  In parallel, nuisance parameters describing contrast loss, readout assignment errors, and relaxation can be included in the likelihood to test self-calibrated sensing.  These additions would turn the present hardware emulation into a genuine metrological validation.

%%%%%%%%%%%%%%%%%%%%%%%%%%%%%%%%%%%%%%%%%%%%%
\section{Conclusions and outlook}
\label{sec:conclusions}

We have reformulated dispersive-interferometric thermometry as a visibility-controlled photon-counting likelihood for the parameters $(\beta,x)$.  The analytical expressions show that the inverse temperature enters through the fringe visibility $\mathcal{V}(\beta)$ and its derivative, while the dispersive interaction enters through the phase of the interference fringe.  This gives a transparent physical interpretation: phase sensitivity is large when the phase-imprinting branch is populated and the fringe contrast is high, whereas temperature sensitivity is appreciable only in the intermediate regime where $|\mathcal{V}'(\beta)|$ is large.

A key operational correction is that a single binary photon-counting setting is rank one for two parameters.  It cannot, by itself, define a locally invertible two-parameter Cram\'er--Rao bound.  Genuine simultaneous estimation requires several controlled settings whose Fisher matrices add to a full-rank $F_{\rm tot}$.  The appropriate scalar cost for such a protocol is then $\mathrm{Tr}[W F_{\rm tot}^{-1}]$, with the weight matrix $W$ fixed by the intended application.  The single-setting Fisher elements remain useful as diagnostic landscapes and as single-parameter bounds when the other parameter is calibrated.

Under amplitude and phase damping, the ideal information is degraded in a strongly probe-dependent manner.  Our comparison of NOON, cat, and squeezed probes should be read as a robustness comparison under matched assumptions, not as a universal ordering of all metrological resources.  NOON probes retain large ideal phase response but are highly loss-sensitive; cat and squeezed probes can preserve useful information over broader operating windows, depending on the noise channel and the target parameter.

The IBM Quantum implementation provides a useful hardware emulation of the $N=1$ likelihood model.  It reproduces the qualitative fringe structure and the corresponding Fisher-information diagnostics, while also showing systematic contrast shrinkage.  This shrinkage biases visibility-based temperature estimates toward $\beta\simeq0$, as expected from an affine-noise model for unmitigated decoherence and state-preparation-and-measurement errors.  We therefore do not present the hardware run as a complete sensing validation.  A complete validation would require blind parameter estimation, empirical error bars for the estimators, and comparison with the full multi-setting Cram\'er--Rao bound as a function of resources.

Overall, visibility provides a useful organizing parameter for dispersive-interferometric sensing, but visibility alone is not a complete multiparameter protocol.  Local identifiability, calibrated noise models, and estimator-level benchmarking determine whether the scheme can be interpreted as a practical joint sensor.

%%%%%%%%%%%%%%%%%%%%%%%%%%%%%%%%%%%%%%%%%

\appendix

\section{Derivation of Effective Visibility under Joint Damping}
\label{app:master_equation}

This appendix derives the effective visibility $\mathcal{V}_{\text{eff}}$ used in Sec.~\ref{subsec:decoherence_models}, starting from the Lindblad master equation that describes the optical mode's evolution under simultaneous amplitude and phase damping.

\subsection{Master equation and coherence evolution}

The evolution of the probe state's density matrix $\hat{\rho}$ is governed by the standard Lindblad master equation \cite{campaioli2024quantum}:
\begin{equation}
    \frac{d\hat{\rho}}{dt} = \gamma_{\text{AD}} \mathcal{D}[\hat{a}](\hat{\rho}) + \gamma_{\text{PD}} \mathcal{D}[\hat{n}](\hat{\rho}),
\end{equation}
where $\gamma_{\text{AD}}$ and $\gamma_{\text{PD}}$ are the amplitude and phase damping rates, respectively, and $\mathcal{D}[\hat{L}](\hat{\rho}) = \hat{L}\hat{\rho}\hat{L}^\dagger - \frac{1}{2}\{\hat{L}^\dagger\hat{L}, \hat{\rho}\}$. Here, $\hat{a}$ is the annihilation operator and $\hat{n} = \hat{a}^\dagger\hat{a}$ is the number operator.

For a coherence element $\rho_{nm}(t) = \langle n|\hat{\rho}(t)|m\rangle$ in the Fock basis, the diagonal ($n=m$) and off-diagonal ($n\neq m$) elements evolve independently. The differential equation for an off-diagonal element is:
\begin{equation}
    \frac{d\rho_{nm}}{dt} = -\left[ \frac{\gamma_{\text{AD}}}{2}(n+m) + \frac{\gamma_{\text{PD}}}{2}(n-m)^2 \right] \rho_{nm}.
\end{equation}

\subsection{Application to NOON-state coherence}

Consider a NOON state $\ket{\psi} = (\ket{N,0} + \ket{0,N})/\sqrt{2}$ in the two-mode representation. After the interferometer, the relevant coherence that determines fringe visibility is the term $\rho_{N0}(t) = \langle N|\hat{\rho}_{\text{out}}(t)|0\rangle$. Setting $n=N$ and $m=0$ in the above equation yields:
\begin{equation}
    \frac{d\rho_{N0}}{dt} = -\left[ \frac{\gamma_{\text{AD}} N}{2} + \frac{\gamma_{\text{PD}} N^2}{2} \right] \rho_{N0}.
\end{equation}
The solution is an exponential decay in time:
\begin{equation}
    \rho_{N0}(t) = \rho_{N0}(0) \exp\left( -\frac{\gamma_{\text{AD}} N t}{2} - \frac{\gamma_{\text{PD}} N^2 t}{2} \right).
\end{equation}

\subsection{Effective visibility with thermal ancilla}

Defining the dimensionless transmission efficiency $\eta = e^{-\gamma_{\text{AD}} t}$ and the dephasing strength $\gamma = \gamma_{\text{PD}} t/2$, the coherence factor becomes $D_N = \eta^{N/2} e^{-\gamma N^2}$.
When the interferometer includes a thermal ancilla with inverse temperature $\beta$, the ideal (unitary) fringe visibility is the operational quantity $\mathcal{V}(\beta)$ defined in Eq.~\eqref{eq:visibility}. The damping channels multiply this ideal visibility by the coherence factor $D_N$, yielding the effective visibility:
\begin{equation}
    \mathcal{V}_{\text{eff}}(\beta,\eta,\gamma) = \eta^{N/2} e^{-\gamma N^2} \mathcal{V}(\beta).
    \label{eq:veff_final}
\end{equation}
This result highlights the distinct scaling behaviors: amplitude damping causes an exponential decay $\propto \eta^{N/2}$, while phase damping causes a super-exponential decay $\propto e^{-\gamma N^2}$. The quadratic scaling with $N$ in the phase-damping exponent underpins the extreme fragility of NOON states to dephasing noise discussed in Sec.~\ref{sec:robustness_analysis}.

Here $\eta^{N/2}$ is the attenuation of the off-diagonal amplitude when the amplitude-damping master equation is applied to the phase-imprinting mode. In the susceptibility analysis below, $(1-\epsilon)^N=\eta^N$ denotes the probability that all $N$ photons survive a photon-loss channel. These are different quantities and should not be identified: the former is an amplitude-level coherence factor for the one-mode damping convention used in Eq.~\eqref{eq:veff_final}, whereas the latter is a full-state survival probability used for the weak-loss scaling estimate.

\section{Fisher-information susceptibility and scaling laws}
\label{app:susceptibility}
\label{app:scalings}

To quantify the \emph{loss-fragility} comparison, we derive the Fisher-information susceptibility ($\chi_F$) for NOON states under local amplitude damping and phase diffusion, and reconcile these local derivatives with the global asymptotic bounds of Escher et al.~\cite{escher2011general}.

\subsection{Definition of susceptibility}
The Fisher-information susceptibility $\chi_F$ measures the leading-order degradation of the quantum Fisher information (QFI) under a weak noise channel $\mathcal{E}_\epsilon$ characterized by a small parameter $\epsilon$ \cite{albarelli2024fisher}:
\begin{equation}
    F(\epsilon) = F(0) - \epsilon \chi_F + O(\epsilon^2).
\end{equation}

\subsection{Susceptibility to amplitude damping (photon loss)}
Consider the NOON state $|\psi\rangle = (|N,0\rangle + |0,N\rangle)/\sqrt{2}$. Under single-photon loss with probability $\epsilon$ (transmission $\eta = 1-\epsilon$), the state evolves into a mixture. Crucially, the loss of a single photon from mode $a$ projects the state to $|N-1,0\rangle$, which is orthogonal to the original superposition. The survival probability of the coherence is $P_{\text{surv}} = (1-\epsilon)^N \approx 1 - N\epsilon$.

The effective visibility scales as $\mathcal{V}_{\text{eff}} \approx 1 - N\epsilon$. Since the QFI for phase estimation scales as $F_Q \propto N^2 \mathcal{V}_{\text{eff}}^2$, expanding to first order yields:
\begin{equation}
    F_Q(\epsilon) \approx N^2 (1 - N\epsilon)^2 \approx N^2 - 2 N^3 \epsilon.
\end{equation}
Identifying terms, we find the susceptibility to amplitude damping scales cubically:
\begin{equation}
    \chi_{\text{AD}} \propto N^3.
\end{equation}
This gives the relative fragility $\chi_F / F(0) \propto N$, which grows linearly with photon number in this weak-loss expansion.

\subsection{Susceptibility to dephasing}
The scaling of dephasing depends on the correlation of the noise.
For \emph{Global Dephasing} (collective fluctuations), the generator $\hat{L} = \hat{n}_a + \hat{n}_b = \hat{N}$ commutes with the NOON state, causing no decoherence.
For \emph{Differential Dephasing} (the relevant metrological noise), the generator is $\hat{L} = \hat{n}_a - \hat{n}_b$. The variance of this generator for a NOON state is $\langle (\Delta \hat{L})^2 \rangle = N^2$. The off-diagonal elements decay as $e^{-\frac{1}{2}\epsilon N^2}$.
Following the expansion $F_Q \propto N^2 (e^{-\frac{1}{2}\epsilon N^2})^2 \approx N^2 (1 - \epsilon N^2)$, we obtain:
\begin{equation}
    F_Q(\epsilon) \approx N^2 - N^4 \epsilon \implies \chi_{\text{PD}} \propto N^4.
\end{equation}
Thus, NOON states are polynomially more fragile to phase diffusion ($\chi \propto N^4$) than to loss ($\chi \propto N^3$). However, in standard hardware, $\epsilon_{\text{AD}}$ often dominates $\epsilon_{\text{PD}}$.

\subsection{Bridging Susceptibility and Asymptotic Bounds}
\label{app:escher_connection}

Finally, we map our local susceptibility metric onto the global asymptotic bounds established by Escher et al.~\cite{escher2011general} and Demkowicz-Dobrza\'{n}ski et al.~\cite{demkowicz2012elusive}.
The Escher bound for lossy interferometry dictates that precision is bounded by the Standard Quantum Limit scaling, $\Delta \phi \geq \frac{\sqrt{1-\eta}}{2\sqrt{\eta N}}$, implying that Heisenberg scaling ($1/N$) is transient.

The susceptibility $\chi_F$ essentially defines the \emph{duration} of this transient regime. We define a \emph{Critical Photon Number} $N_{\text{crit}}$ where the noise-induced reduction becomes comparable to the initial quantum advantage:
\begin{equation}
    N^2 \sim \epsilon \cdot \chi_F(N).
\end{equation}
Substituting our derived susceptibility for loss, $\chi_{\text{AD}} \propto N^3$, we recover:
\begin{equation}
    N^2 \sim \epsilon N^3 \implies N_{\text{crit}} \sim \frac{1}{\epsilon}.
\end{equation}
This recovers the expected order-of-magnitude breakdown point of NOON states. For $N < N_{\text{crit}}$, the state operates in the \emph{loss-fragility} regime where quantum advantage is possible. For $N > N_{\text{crit}}$, the cubic susceptibility overwhelms the quadratic gain.
In contrast, squeezed vacuum states have a susceptibility structure consistent with the lossy-interferometry bounds and can maintain a constant-factor advantage over the SQL under suitable measurement conditions.  This probe-family robustness comparison can therefore be interpreted as a ranking of the useful transient window for different probe architectures.

\section{Theoretical Model of Experimental Bias}
\label{app:bias_model}

\begin{figure}[ht]
    \centering
    \includegraphics[width=0.95\columnwidth]{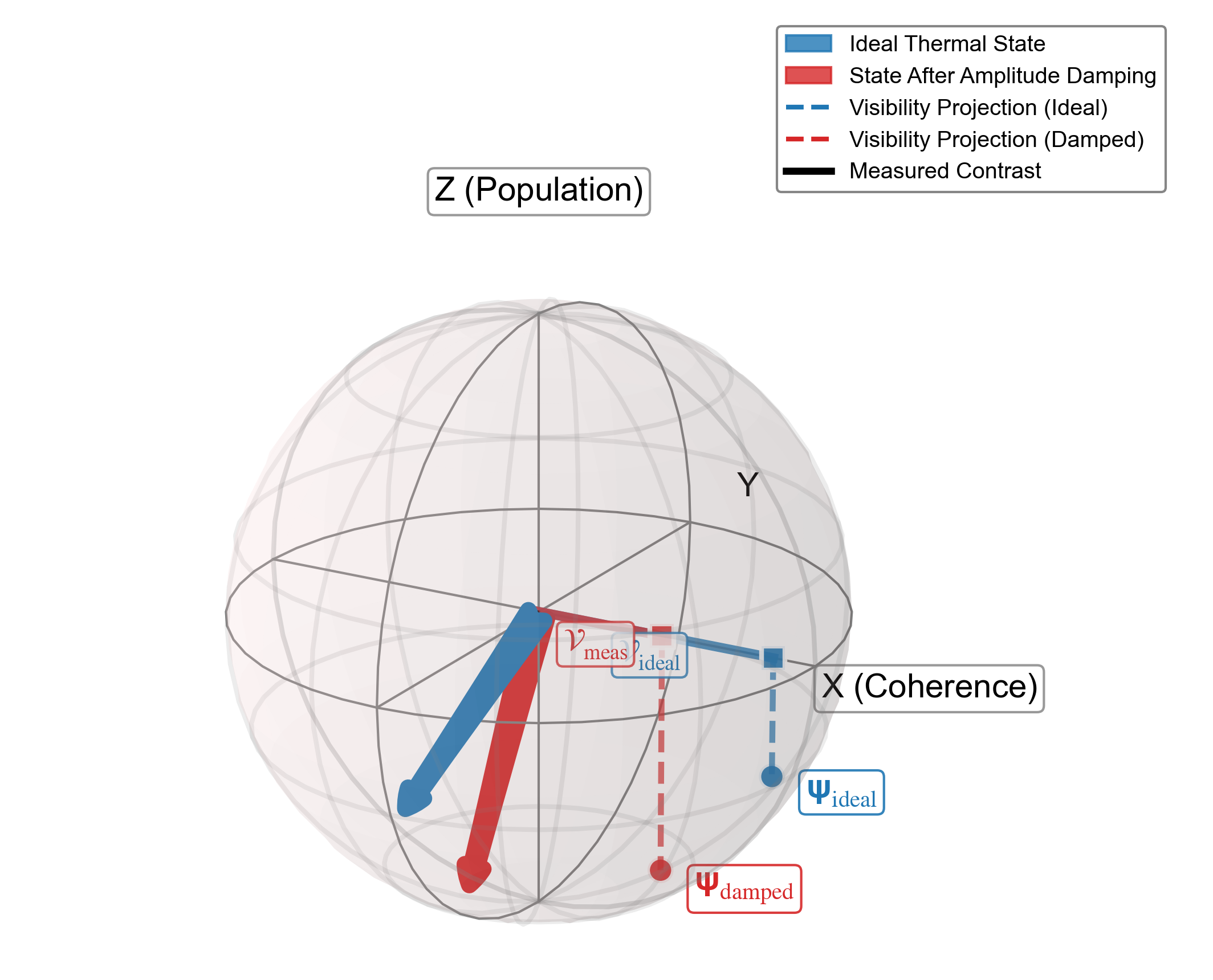}
    \caption{Schematic visualization of a contrast-shrinkage bias mechanism. Unmitigated noise channels and SPAM errors effectively reduce the observed fringe contrast, pulling the inferred visibility toward a baseline close to $\mathcal{V}(0)=1/3$. When $\beta$ is estimated by inverting the monotonic map $\mathcal{V}(\beta)$, this shrinkage biases $\hat{\beta}$ toward $\beta\simeq 0$, yielding $|\hat{\beta}|<|\beta_{\rm true}|$ over broad operating regions.}
    \label{fig:bloch_bias}
\end{figure}

The contrast-shrinkage mechanism analyzed below is summarized schematically in Fig.~\ref{fig:bloch_bias}.

This appendix provides an analytical model for the systematic bias observed in the experimental temperature estimates (Sec.~\ref{sec:experiment_ibm}), where $\hat{\beta}_{\text{exp}}$ underestimates the true magnitude $|\beta_{\text{true}}|$.
In the present framework, $\beta$ is inferred operationally through the fringe visibility $\mathcal{V}(\beta)$ defined from the output probabilities.

\subsection{Estimator construction and inverse map}

From Eq.~\eqref{eq:visibility}, the visibility is
\begin{equation}
  \mathcal{V}(\beta)=\frac{e^{-\beta\hbar\omega_0}}{2+e^{-\beta\hbar\omega_0}},
  \qquad 0<\mathcal{V}<1.
\end{equation}
Inverting this monotonic map yields the estimator
\begin{equation}
  \hat{\beta}
  = \mathcal{V}^{-1}(\mathcal{V}_{\rm meas})
  = -\frac{1}{\hbar\omega_0}\,
    \ln\!\left(\frac{2\,\mathcal{V}_{\rm meas}}{1-\mathcal{V}_{\rm meas}}\right).
  \label{eq:beta_from_V}
\end{equation}
Operationally, $\mathcal{V}_{\rm meas}$ can be obtained either directly from the measured extrema of $P_0(\beta,x)$ (e.g., via $P_{\max}$ and $P_{\min}$) or by fitting a fringe model to $P_0$ versus $x$ at fixed $\beta$.

\subsection{Contrast shrinkage and bias toward \texorpdfstring{$\beta\simeq 0$}{beta approximately 0}}

A minimal phenomenological description of unmitigated decoherence and SPAM errors is that they partially randomize the output statistics and pull the inferred visibility toward a baseline.
Since $\mathcal{V}(0)=1/3$ plays the role of the infinite-temperature reference point in the ideal model, we model
\begin{equation}
  \mathcal{V}_{\rm meas}
  \simeq \kappa\,\mathcal{V}(\beta_{\rm true}) + (1-\kappa)\,\mathcal{V}(0),
  \qquad 0<\kappa<1.
  \label{eq:V_shrink}
\end{equation}

This affine shrinkage has two key consequences:
(i) if $\beta_{\rm true}>0$ then $\mathcal{V}(\beta_{\rm true})<1/3$, so Eq.~\eqref{eq:V_shrink} increases the observed visibility;
(ii) if $\beta_{\rm true}<0$ then $\mathcal{V}(\beta_{\rm true})>1/3$, so Eq.~\eqref{eq:V_shrink} decreases it.
In both cases, $\mathcal{V}_{\rm meas}$ is moved closer to $1/3$, and therefore $\hat{\beta}$ obtained from Eq.~\eqref{eq:beta_from_V} is biased toward $0$, i.e.,
\begin{equation}
  |\hat{\beta}|\;<\;|\beta_{\rm true}|
\end{equation}
over a broad range of parameters (consistent with the trends in Table~\ref{tab:exp_estimates}).

\subsection{Correction protocol}

If $\kappa$ can be independently calibrated (e.g., by preparing reference points with known $\beta$ or by dedicated process tomography), then an unbiased estimate can be obtained by undoing the shrinkage:
\begin{equation}
  \hat{\beta}_{\rm corr}
  = \mathcal{V}^{-1}\!\left(\frac{\mathcal{V}_{\rm meas}-(1-\kappa)\mathcal{V}(0)}{\kappa}\right),
\end{equation}
provided the argument lies in $(0,1)$.
Alternatively, in a multiparameter inference one may treat $\kappa$ as a nuisance parameter and jointly fit $(\beta,\kappa)$ from the same fringe data, trading additional statistical uncertainty for immunity to systematic bias.

\section{Comparative probe-family robustness}
\label{app:hierarchy_derivation}

\begin{table*}[t]
\caption{Probe-family robustness comparison of quantum probe states. The susceptibility $\chi_F$ quantifies the linear degradation of QFI under weak amplitude damping, while the robustness index $\mathcal{R}$ measures the preserved fraction at $\epsilon=0.1$.}
\label{tab:hierarchy_summary}
\centering
\begin{tabular}{lcccc}
\hline
\textbf{Probe State} & \textbf{Ideal QFI} ($F_0$) & \textbf{Susceptibility} ($\chi_F$) & \textbf{Scaling} & \textbf{Robustness} $\mathcal{R}(\epsilon=0.1)$ \\
\hline
NOON State & $\sim N^2$ & $\sim N^3$ & Exponential collapse & $< 0.01$ (Fragile) \\
Cat State & $\sim \bar{n}^2$ & $\sim \bar{n}^3$ & Exponential decay & $\sim e^{-0.4\bar{n}}$ (Conditional) \\
squeezed vacuum & $\sim \bar{n}^2$ (lossless) & $\sim \bar{n}$ & Graceful degradation & $\sim 0.9$ (gradual loss) \\
& $\sim \bar{n}$ (lossy) & & & \\
\hline
\end{tabular}
\end{table*}

This appendix summarizes the analytical basis for the probe-family robustness comparison presented in Fig.~\ref{fig:decoherence_landscape}, extending the susceptibility analysis of Appendix~\ref{app:susceptibility} to Schr\"{o}dinger cat and squeezed vacuum states. We quantify their relative robustness under amplitude damping and explain their distinct operational regimes.

\subsection{Schr\"{o}dinger cat states: non-Gaussian resilience}

Consider the even-parity cat state $|\psi_{\text{cat}}\rangle = \mathcal{N}(|\alpha\rangle + |-\alpha\rangle)$, where $\mathcal{N} = [2(1+e^{-2|\alpha|^2})]^{-1/2}$ and the mean photon number is $\bar{n} \approx |\alpha|^2$ for $|\alpha| \gg 1$. Unlike NOON states, photon loss does not immediately destroy quantum coherence but rather induces a parity flip: $\hat{a}|\text{even}\rangle \to \alpha|\text{odd}\rangle$. This can preserve phase-space structure when parity information is tracked, rather than immediately projecting onto the original even-cat manifold.

The off-diagonal coherence element $\langle\alpha|-\alpha\rangle = e^{-2|\alpha|^2}$ decays under amplitude damping with transmission $\eta$ as:
\begin{equation}
    \mathcal{C}_{\text{cat}}(\eta) \approx \exp\left[-2|\alpha|^2(1-\sqrt{\eta})\right].
\end{equation}
For weak loss ($\eta = 1-\epsilon$, $\epsilon \ll 1$), this becomes $\mathcal{C}_{\text{cat}} \approx e^{-2\bar{n}\epsilon}$. The quantum Fisher information for parameter $\theta$ encoded via a unitary $\hat{U}_\theta = e^{i\theta\hat{n}}$ consequently scales as:
\begin{equation}
    F_{\text{cat}}(\theta) \approx 4\bar{n}^2 e^{-4\bar{n}\epsilon}.
\end{equation}
The Fisher-information susceptibility therefore scales as:
\begin{equation}
    \chi_{F}^{\text{(cat)}} \propto \bar{n}^3 e^{-4\bar{n}\epsilon} \bigg|_{\epsilon\to 0} \sim \bar{n}^3.
\end{equation}

Despite this cubic scaling, cat states maintain Heisenberg-limited sensitivity ($F \propto \bar{n}^2$) until $\epsilon \sim 1/\bar{n}$. This contrasts sharply with NOON states, which fail catastrophically at $\epsilon \sim 1/N$. The critical advantage is that cat states retain phase sensitivity even after multiple photon losses, provided the parity-flipped state remains coherent. This makes them relevant candidates for transient thermometry protocols where non-Markovian information backflow is available \cite{mendonca2025information}.

\subsection{Squeezed vacuum states: Gaussian robustness}

The two-mode squeezed vacuum (TMSV) state $|\xi\rangle = \hat{S}_{12}(\xi)|0,0\rangle$, with squeezing parameter $r$ ($\xi = re^{i\phi}$) and mean photon number per mode $\bar{n} = \sinh^2 r$, is a standard Gaussian probe family for lossy interferometric sensing. Under amplitude damping, the state remains Gaussian, with the covariance matrix transforming linearly. For phase estimation in a lossy Mach--Zehnder interferometer with transmission $\eta$, we use the following standard Gaussian-probe benchmark expression under the same phase convention $2x=\chi t$; convention-dependent factors only rescale $F_{xx}$ and do not affect the robustness trends discussed here \cite{escher2011general}:
\begin{equation}
    F_{\text{SQ}}(x) = \frac{\eta(\bar{n}^2 + \bar{n})}{1-\eta + \frac{1}{2\bar{n}+1}}.
\end{equation}
In the high-squeezing limit ($\bar{n} \gg 1$), this simplifies to:
\begin{equation}
    F_{\text{SQ}}(x) \approx \frac{\eta \bar{n}}{1-\eta} \quad \text{for fixed } \eta < 1.
\end{equation}
This highlights two relevant features:
\begin{enumerate}
    \item For any finite loss ($\eta < 1$), the scaling transitions from Heisenberg ($\bar{n}^2$) to standard quantum limit ($\bar{n}$) as $\bar{n}$ increases.
    \item The degradation is \emph{gradual}---there is no catastrophic threshold as with NOON states.
\end{enumerate}

The susceptibility for squeezed states scales linearly with $\bar{n}$:
\begin{equation}
    \chi_{F}^{\text{(SQ)}} \propto \bar{n}.
\end{equation}
This linear scaling, combined with the SQL scaling of $F_{\text{SQ}}$, means the \emph{relative fragility} $\chi_F/F$ remains constant, unlike for NOON states where it grows with $N$. This scaling explains their gradual degradation under loss and their usefulness for steady-state sensing in lossy environments.

\subsection{Temperature-sensitivity considerations}

For temperature estimation, the comparison changes. Cat states encode temperature information in the \emph{parity} of the interference pattern, which is more sensitive to dephasing than the phase information utilized by squeezed states. Consequently:
\begin{itemize}
    \item \textbf{NOON states}: Highly sensitive in the ideal model but fragile to loss and dephasing.
    \item \textbf{Cat states}: More tolerant to amplitude damping than NOON states in the plotted regime, but still vulnerable to phase damping for temperature estimation.
    \item \textbf{Squeezed states}: Robust for phase estimation under loss, but their temperature performance depends on the available measurement strategy.
\end{itemize}

\subsection{Summary of the probe-family comparison}

Table~\ref{tab:hierarchy_summary} summarizes these analytical results, providing a representative comparison of the three probe classes under the assumptions stated in Sec.~\ref{subsec:probe_family_robustness}. The \emph{Robustness Index} $\mathcal{R} = F_{\text{lossy}}/F_{\text{ideal}}$ at fixed $\epsilon = 0.1$ illustrates the practical performance degradation.

This comparison helps explain the trade-offs discussed in Sec.~\ref{sec:robustness_analysis}: NOON states provide high peak sensitivity but are strongly loss-sensitive; cat states can be useful in transient non-Gaussian regimes; and squeezed vacuum states are attractive for steady-state sensing under lossy conditions when compatible measurements are available.

\begin{acknowledgments}
 We acknowledge financial support from the Brazilian agencies: Coordena\c{c}\~{a}o de Aperfei\c{c}oamento de Pessoal de N\'{i}vel Superior (CAPES), Finance Code 001. NGA thanks FAPESP Grant 2024/21707-0.
\end{acknowledgments}
\subsection*{Artificial intelligence usage declaration}
In accordance with journal guidelines, the authors disclose that AI-assisted tools were used for language editing and assistance with literature searches. All text, references, calculations, interpretations, and visual content were independently reviewed and verified by the authors, who assume full responsibility for the integrity, accuracy, and originality of the manuscript. No AI tool fulfills the role of, or is listed as, an author.

\bibliographystyle{apsrev4-2}
\bibliography{manuscript}

\end{document}